\documentclass[a4paper,twocolumn,11pt,accepted=2025-06-30]{quantumarticle}
\pdfoutput=1
\usepackage{amsmath,amssymb,amsfonts}
\usepackage{amsthm, blkarray}
\usepackage{mathtools}
\usepackage{hyperref}
\usepackage{float}
\usepackage{graphicx}
\usepackage{enumerate}
\usepackage[dvipsnames]{xcolor}
\usepackage{physics}
\usepackage[T1]{fontenc}
\usepackage{xr}
\usepackage[dvipsnames]{xcolor}
\usepackage{bbold}

\usepackage{tikz}
\usetikzlibrary{quantikz}

\newcommand{\id}{\mathbb{1}}
\newcommand{\upe}{\mathrm{e}}
\newcommand{\upi}{\mathrm{i}}
\newcommand{\upd}{\mathrm{d}}

\newcommand{\crot}{\textsc{crot}}
\newcommand{\cz}{\textsc{cz}}

\newcommand{\cA}{\mathcal{A}}

\newcommand{\cD}{\mathcal{D}}
\newcommand{\cE}{\mathcal{E}}
\newcommand{\cF}{\mathcal{F}}

\newcommand{\cI}{\mathcal{I}}

\newcommand{\cN}{\mathcal{N}}

\begin{document}
\title{Circuit-level fault tolerance of cat codes}
\author{Long D. H. My}
\affiliation{Yale-NUS College, Singapore}
\affiliation{Centre for Quantum Technologies, National University of Singapore{, Singapore}}
\orcid{0000-0002-4628-6492}
\author{Shushen Qin}
\affiliation{Centre for Quantum Technologies, National University of Singapore{, Singapore}}
\orcid{0009-0000-7872-9393}
\author{Hui Khoon Ng}
\email{huikhoon.ng@nus.edu.sg}
\affiliation{Yale-NUS College, Singapore}
\affiliation{Centre for Quantum Technologies, National University of Singapore{, Singapore}}
\orcid{0000-0003-2397-840X}

\maketitle

\begin{abstract}
Bosonic codes encode quantum information into a single infinite-dimensional physical system endowed with error correction capabilities. This reduces the need for complex management of many physical constituents compared with standard approaches employing multiple physical qubits. Recent discussions of bosonic codes centre around correcting only boson-loss errors, with phase errors either actively suppressed or deferred to subsequent layers of encoding with standard qubit codes. Rotationally symmetric bosonic (RSB) codes, which include the well-known cat and binomial codes, are capable of simultaneous correction of loss and phase errors, offering an alternate route that deals with arbitrary errors already at the base layer. Here, we investigate the robustness of such codes, moving away from the more idealistic past studies towards a circuit-level noise analysis closer to the practical situation where every physical component in the device is potentially faulty. We extend the concept of fault tolerance to the case of RSB codes, and then examine the performance of two known error correction circuits under circuit-level noise. Our analysis reveals a significantly more stringent noise threshold for fault-tolerant operation than found in past works; nevertheless, we show how, through waiting-time optimization and the use of squeezing, we can restore the noise requirements to a regime achievable with near-term quantum hardware. While our focus here is on cat codes for concreteness, a similar analysis applies for general RSB codes.
\end{abstract}

\section{Introduction}
\label{sec:intro}

The realisation of quantum computers faces significant challenges due to noise from unwanted couplings to the environment and control imperfections. To build useful large-scale quantum computers, quantum error correction and associated fault-tolerant procedures will be necessary. Most fault-tolerant quantum computing schemes are based on encoding each logical qubit onto multiple physical qubits (see, for example, Refs.~\cite{shor1996fault, steane1997active,gottesman1998theory, knill2005quantum, fowler2012surface}), with recent experimental advances showing impressive progress in bringing the device noise closer to critical threshold values, below which error correction can offer genuine advantage \cite{egan2021fault, google2021exponential, krinner2022realizing, zhao2022realization, google2023suppressing, bluvstein2024logical}. However, these schemes demand significant overheads, in terms of physical-qubit numbers, to attain the scale and accuracy when quantum computers become useful \cite{chamberland2017overhead, kim2022fault, beverland2022assessing}. 
Bosonic codes, however, offer the possibility of encoding a logical qubit onto a \emph{single} infinite-dimensional physical system (e.g., a harmonic oscillator), while offering protection against errors that arise in such systems \cite{chuang1997bosonic,cochrane1999macroscopically,gottesman2001encoding,michael2016new}. Recent experiments with bosonic codes demonstrated lifetimes of error-corrected qubits that can exceed that of unencoded qubits realized using the same hardware \cite{ofek2016extending,hu2019quantum,sivak2023real}. This makes bosonic qubits an intriguing avenue for near-term hardware, especially when the number of physical components remain practically limited \cite{noh2022low,chamberland2022building}. 

Currently, much of the bosonic-codes discussion centres around correcting losses, the inherent and dominant source of noise in an oscillator, deferring the suppression of phase errors to subsequent layers of encoding \cite{ofek2016extending,li2017cat,albert2018performance}, or engineering highly biased-noise qubits \cite{mirrahimi2014dynamically,leghtas2015confining,lescanne2020exponential,puri2017engineering,grimm2020stabilization,gautier2022combined}. Phase errors, however, naturally arise when the oscillator is coupled to auxiliary systems for universal control \cite{ofek2016extending, rosenblum2018fault} and this becomes the limiting factor when doing gates. While bias-preserving gates can be used to maintain the noise bias \cite{puri2020bias, gautier2022combined, xu2022engineering}, such approaches limit the variety of gates that can be done easily. An alternative approach is to use bosonic codes capable of correcting loss and phase errors simultaneously. Rotationally symmetric bosonic (RSB) codes \cite{grimsmo2020quantum} form a large family of such codes.

Examples of RSB codes include known bosonic codes like the cat codes \cite{cochrane1999macroscopically, mirrahimi2014dynamically} and the binomial codes \cite{michael2016new}. Ref.~\cite{grimsmo2020quantum} unified these examples to give a general theory of RSB codes, and constructed error-correction (EC) circuits for a general RSB code, showing how loss and phase errors on the input to the EC circuit can be corrected. The authors demonstrated a code-capacity break-even point---the noise level below which the encoded qubits have less errors than unecoded ones---as high as a 5\% probability of error. Subsequently, Ref.~\cite{hillmann2022performance} extended these ideal-EC results to incorporate the impact of measurement imperfections, replacing the difficult-to-realise canonical phase measurement used in Ref.~\cite{grimsmo2020quantum} by implementable ones and considering finite detection efficiencies. That study revealed the significant impact of detection efficiency. Nevertheless, even with an efficiency as low as $50\%$, RSB codes are still able to correct loss and phase errors, though the break-even point now drops to $\sim 0.1$--$1\%$.

These past analyses, however, tell only part of the story of how RSB codes perform on real devices. In actual operation, faults can occur anywhere in the EC circuit, not just in the inputs or the measurements, causing errors that can spread and reduce the correction capacity of the underlying codes. A full fault-tolerance analysis has to account for faults occurring in individual circuit components. In particular, the EC circuits of Ref.~\cite{grimsmo2020quantum} involve a novel two-mode gate that may be very noisy at least in near-term devices. Only a circuit-level analysis can provide realistic estimates of the noise requirements for RSB codes to function well. 

Our work gives such a circuit-level fault-tolerance analysis for RSB codes. For concreteness, we focus on cat codes \cite{cochrane1999macroscopically,mirrahimi2014dynamically,leghtas2013hardware} given their current experimental popularity; our approach nevertheless extends easily to other RSB codes.

\subsection{Summary of results}
We begin by extending the concepts of standard fault tolerance, originally applied to qubit codes, to the situation of RSB codes with discrete loss errors but continuous phase errors. Formally, for fault tolerance, a small number of loss errors and phase errors of a small magnitude at any circuit location cannot lead to a logical error in the output of the circuit. Furthermore, a faulty EC circuit---or ``gadget"---still has to function sufficiently well so that the output errors depend only on the faults that occurred in the gadget, not on the input errors; such a property ensures that errors from consecutive EC gadgets do not accumulate into uncorrectable ones that can cause the whole computation to fail.

We then apply our notion of fault tolerance for RSB codes to investigate the performance of two EC gadgets introduced in Ref.~\cite{grimsmo2020quantum}, under circuit-level noise where all circuit components are potentially faulty. We show that both EC gadgets are fault tolerant in the formal sense, provided the noise is below some (pseudo)threshold to allow for successful error removal even with faulty components. We numerically obtain the break-even points for a memory task, where we interleave EC gadgets with memory (waiting) steps, applying circuit-level noise throughout. Our simulations show that the circuit-level noise thresholds (see Fig.~\ref{fig:fixed_wait_threshold}) fall short by one to two orders of magnitude when compared with earlier more optimistic $10^{-2}$ predictions in the ideal-EC scenario; this is so even after optimizing the correction capability of the code over the amplitude of the cat code. This directly highlights the importance of a circuit-level analysis for judging the efficacy of a fault-tolerance scheme.

The circuit-level noise thresholds can nevertheless be improved to levels feasible with near-term hardware. We show that, by optimizing the waiting time between EC cycles and the use of squeezed cat codes, error probabilities of about $0.1\%$, within reach of experiments, can already be below the break-even boundary; see Fig.~\ref{fig:squeeze_threshold}.

In all our numerical studies, we observe a notable difference in performance between the two EC schemes of Ref.~\cite{grimsmo2020quantum}, stemming from differences in robustness against noise in the ancillary mode, a difference not at all apparent in the ideal-EC analysis of the original paper. Our fault-tolerance analysis further reveals how one should choose the codes for the ancillary modes in the two schemes, to maximize robustness against errors.

\subsection{Organization of the article}
The rest of the article is structured as follows. In Sec.~\ref{sec:prelim}, we establish some basic concepts needed for our work, including the cat codes, and the two EC gadgets introduced in Ref.~\cite{grimsmo2020quantum}. In Sec.~\ref{sec:ft_analysis}, we explain the fault tolerance property relevant for the RSB-codes situation and show that the EC gadgets are fault tolerant under circuit-level noise that is not too strong. In Sec.~\ref{sec:num_sim}, we present our numerical results for the performance of EC gadgets in the presence of a concrete circuit-level noise model, for the scenarios of fixed and variable waiting periods, and with the use of squeezed cat codes. We conclude in Sec.~\ref{sec:concl} and offer some outlook on the steps forward.

\section{Preliminaries}
\label{sec:prelim}
We begin with a few preliminary concepts, following the description in Ref.~\cite{grimsmo2020quantum}, but specializing to the cat-code situation, the focus of our paper.

\subsection{The error model}
\label{sec:noise_model}

To describe errors that can occur in a bosonic mode, we begin with the operator basis used in Ref.~\cite{grimsmo2020quantum},
\begin{equation}
	\{\upe^{\upi\theta\widehat n}\widehat a^k, (\widehat a^\dagger)^k\upe^{-\upi\theta\widehat n}\},
\end{equation}
with $k$ a nonnegative integer, and $\theta\in[0,2\pi)$. Here, $\widehat a$ and $\widehat a^\dagger$ are the bosonic annihilation and creation operators with $[\widehat a,\widehat a^\dagger]=1$, and $\widehat n\equiv \widehat a^\dagger \widehat a$ is the number operator. Of particular relevance for the noise seen in current devices are the boson-loss terms associated with $\widehat a^k$, and phase-space rotation, $R(\theta)\equiv\upe^{\upi\theta\widehat n}$. We refer to $\widehat a^k$ as an occurrence of $k$ loss errors, and $R(\theta)$ as a phase (or rotation) error parameterized by a continuous angle $\theta$. A general error can be a combination of the two: $R(\theta)\widehat a^k$. Boson-gain terms $(\widehat a^\dagger)^k$ can also be of relevance (e.g., from thermal excitation), but they are often observed to be of secondary importance in current devices. We will hence focus only on correcting loss and phase errors.

We note that a noisy gate, where errors can occur \emph{during} its operation, can always be written as an ideal gate followed by an error channel. This uses a standard argument employed in noise analyses where we write, without loss of generality, the noisy gate $\widetilde U$ as $\widetilde U = \mathcal{E}\circ U$ (as a composition of quantum channels), with $\mathcal{E}\equiv \widetilde U\circ U^\dagger$ being the error (or noise) channel and $U$ is the noise-free gate operation. $\cE$ can then be written in terms of the operator basis above.

\subsection{Cat codes}
\label{sec:cat_code}

Cat codes form one of the most studied and experimentally feasible classes of RSB codes. RSB codes are single-mode bosonic codes that support a single logical qubit. An order-$N$ RSB code is defined by the requirement that, for $N=1,2,3,\ldots$, the code space is invariant under the phase-space rotation $R_N \equiv R(\frac{2\pi}{N})= \upe^{\upi \tfrac{2\pi}{N} \widehat{n}}$, and that $Z_N\equiv \sqrt{R_N}=R_{2N}$ acts as the Pauli-$Z$ operator for the logical qubit. One can describe the code using a primitive state $\ket{\Theta}$, such that the $\ket{0_N}$ and $\ket{1_N}$ code states---the $\pm1$-eigenstates of $Z_N$---are written as
\begin{align}\label{eq:codeword_z_rot}
	\ket{\mu_N} &= \frac{1}{\sqrt{\cN_{\mu}}} \sum_{m=0}^{2N-1} (-1)^{\mu m} {Z_N}^m \ket{\Theta},\quad \textrm{for}~\mu=0,1,
\end{align}
where $\cN_{\mu}$ are normalization constants. The code space is the linear span of $\ket{0_N}$ and $\ket{1_N}$.

Cat codes are RSB codes with the primitive state $\ket{\Theta}= \ket\alpha$, a coherent state with real $\alpha$. The $\ket{0_N}$ and $\ket{1_N}$ states of a cat code are superpositions of (complex-parameter) coherent states, since $R_k\ket{\alpha}=\big|\upe^{\upi\frac{2\pi}{k}}\alpha\big\rangle$:
\begin{align}\label{eq:cat_codeword}
	\ket{\mu_{N, \alpha}} = \frac{1}{\sqrt{\cN_{\mu}}}\sum_{m=0}^{2N-1} (-1)^{\mu m} \ket{e^{\upi \tfrac{\pi m}{N}}\alpha},\quad \mu=0,1.
\end{align}
Here, the $\alpha$ subscript in $\ket{\mu_{N,\alpha}}$ indicates that these are code states of a cat code.
In the Fock basis, these states take the form
\begin{align}
	\label{eq:cat_code_fock}
	\ket{\mu_{N, \alpha}} &= \frac{2N\upe^{-\alpha^2/2}}{\sqrt{\cN_\mu}}\sum_{m\in\mathbb{Z}_\mu} \frac{\alpha^{mN}}{\sqrt{(mN)!}} \ket{mN},
\end{align}
where we adopt the shorthand $\mathbb{Z}_0\equiv \{0,2,4,\ldots\}$ for the set of even nonnegative integers, while $\mathbb{Z}_1\equiv \{1,3,5,\ldots\}$ is the set of odd nonnegative integers. 
We note that the normalization constants can be written as $\cN_{\mu} = 2N + O(\upe^{-c\alpha^2})$ with $c = 1 - \cos(\frac{\pi}{N})$, so that $\cN_0 \approx \cN_1 \approx 2N$ when $\alpha\rightarrow\infty$. In this limit, the conjugate-basis code states can be written approximately as superpositions of coherent states with even- and odd-multiple-$\pi/N$ phases,
\begin{equation}\label{eq:approx_x_codewords}
	\ket{\pm_{N,\alpha}}\equiv \frac{1}{\sqrt 2}(\ket{0_{N,\alpha}}\pm\ket{1_{N,\alpha}})\approx\frac{1}{\sqrt N}\!\!\sum_{\ell\in\mathbb{Z}_{N,\pm}}\!\!\ket{\alpha\upe^{\upi\frac{\ell\pi}{N}\!\!}},
\end{equation}
with the additional shorthand $\mathbb{Z}_{N,+}\equiv \{0,2,4,\ldots,2N-2\}$ and $\mathbb{Z}_{N,-}\equiv \{1,3,5,\ldots,2N-1\}$.

The error-correcting properties of the cat (and of the general RSB) codes can be understood intuitively. As $\ket{0_N}$ is supported only on even multiple-$N$ states $\{\ket{kN}:k\in\mathbb{Z}_0\}$, while $\ket{1_N}$ is supported only on odd multiple-$N$ states $\{\ket{kN}:k\in\mathbb{Z}_1\}$, there is a ``distance'' $N$ between the two states such that we need $N$ losses (i.e., $\widehat a^N$) to go from one state to the other, for a logical bit-flip. For phase errors, we note that the conjugate-basis code states are separated by an angle $\pi/N$ in phase space, i.e., $R_{2N}\ket{\pm_N}=\ket{\mp_N}$. Phase errors with angles smaller than $\pi/N$ are hence detectable in this conjugate basis.

Concretely, from the Knill-Laflamme conditions for error correction (see App.~\ref{app:knill_laflamme}), an order-$N$ cat code can approximately correct up to $N-1$ loss errors and a continuous range of phase errors of total angle less than $\frac{\pi}{N}$. More precisely, the set of simultaneously correctable errors can be chosen to be
\begin{align}\label{eq:QEC_E}
	&\cE \equiv \{R(\theta)\widehat{a}^k\},
\end{align}   
with $k\textrm{ nonnegative integer}\in[k_0,k_0+N)$ and $\theta \textrm{ real}\in(\theta_0,\theta_0+\tfrac{\pi}{N}]$.
We note the trade-off between correcting for loss versus phase errors: A larger-$N$ cat code corrects more loss errors but a smaller range of phase errors. That the cat code \emph{approximately} corrects loss and phase errors comes from the violation of the Knill-Laflamme conditions by a factor $\upe^{-c'\alpha^2}$ for some constant $c'$ (see App.~\ref{app:knill_laflamme}), vanishingly small as $\alpha$ gets large. 

We will use a ``bar'' to denote a logical operator---e.g., $\overline Z$ for the logical Pauli-$Z$ operator, equal to $Z_N$ for an order-$N$ cat code---whenever we want to avoid specifying the code choice. This is particularly useful when there are multiple modes encoded in different codes.

\subsection{$\crot$ gates}
\label{sec:logical_ops}
The EC gadgets we will study below employ a one-parameter family of two-mode gates referred to in Ref.~\cite{grimsmo2020quantum} as the controlled-rotation gates $\crot(\varphi)$,
\begin{equation}\label{eq:crot}
	\crot(\varphi) \equiv \upe^{\upi \varphi \widehat{n}\otimes \widehat{n}},
\end{equation}
for $\varphi$ a real number. Since they are defined on the physical modes, the $\crot$ gates are well-defined operations on \emph{any} pair of RSB codes, including ones with different orders and different primitive states. Of particular interest is the case of $\varphi=\frac{\pi}{NM}$ for two RSB codes, one of order $N$, the other of order $M$ (possibly with different primitive states). One can check that $\crot{\left(\frac{\pi}{MN}\right)}$ acts as the logical controlled-$Z$ ($\overline\cz$) gate between the two code spaces. Note that, because of the symmetry between the two modes in the definition of $\crot$, we need not specify which mode is the control and which is the target.

\begin{figure}
	\begin{quantikz}[row sep=0cm, column sep = 0.3cm]
		\lstick{$\widehat{a}^k$}&\gate[wires=2]{\crot(\varphi)}&\qw\rstick{$\widehat{a}^k$}\\
		\qw&\qw&\qw\rstick{$R(-k\varphi)$}
	\end{quantikz}
	\begin{quantikz}[row sep=0cm, column sep = 0.3cm]
		\lstick{$R(\theta)$}&\gate[wires=2]{\crot(\varphi)}&\qw\rstick{$R(\theta)$}\\
		\qw&\qw&\qw
	\end{quantikz}
	\caption{\label{fig:errorCROT}Circuit identities for the propagation of loss ($\widehat a$; we consider $k$ of them) and phase [$R(\theta)\equiv \upe^{\upi\theta\widehat n}$] errors through a $\crot$ gate on two bosonic modes.} 
\end{figure}
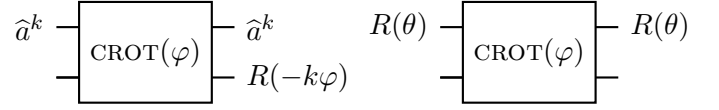

Two operator identities---straightforward to verify---for how a $\crot$ gate propagates loss and phase errors will be of relevance in our discussion below:
\begin{align}\label{eq:errorCROT}
	\crot(\varphi)(\widehat{a}^k\otimes\id)&=[\widehat{a}^k\otimes R(-k\varphi)]\crot(\varphi)\nonumber\\
	\textrm{and}\quad\crot(\varphi)[R(\theta)\otimes\id]&=[R(\theta)\otimes\id]\crot(\varphi).
\end{align}
These are more easily visualized in a circuit representation; see Fig.~\ref{fig:errorCROT}.

\subsection{Measurements}
\label{sec:meas}
Next, we introduce the measurements needed in the EC gadgets. First is the canonical phase measurement; we need only the discrete version, but we begin with the general case. The canonical phase measurement refers to the optimal measurement for the one-parameter estimation problem on bosonic systems \cite{holevo2011probabilistic}: What is $\theta$ in the state $R(\theta)\ket{\Phi_0}$, for a known state $\ket{\Phi_0}$? In our case, we set $\ket{\Phi_0}\equiv\ket{+_{N,\alpha}}$ for an order-$N$ cat code. The canonical phase measurement can then be written as the one-parameter family of measurement (or POVM) operators,
\begin{equation}
	\Pi(\phi)\equiv \ket\phi\bra\phi,\quad\textrm{with }\ket\phi\equiv \frac{1}{\sqrt{2\pi}}\sum_{n=0}^\infty \upe^{\upi n\phi}\ket n,
\end{equation}
for $\phi$ going over any given $2\pi$ range, and with the completeness relation $\int_{(2\pi)}\upd\phi ~\Pi(\phi)=\id$; $\{\ket n\}$ is the Fock basis. The $\ket\phi$ are the so-called ``phase states" \cite{holevo2011probabilistic}, which are in fact un-normalizable, rendering this canonical phase measurement difficult to implement in practice. Nevertheless, there are proposals to approximate it using heterodyne or adaptive homodyne measurements \cite{wiseman1998adaptive, blais2021circuit}. Reference \cite{hillmann2022performance} showed that the uncertainty of the adaptive homodyne measurement differs from that of the canonical phase measurement by only a negligible amount. We thus continue to work with the canonical phase measurement, assuming that its implementable approximation does not significantly degrade the measurement accuracy. 

In the EC gadgets, we need only the discrete version of the canonical phase measurement. Specifically, we are interested in the discrete estimation problem: What is the integer $k\in\{0,1,2,\ldots,K\}$ (for $K$ a positive integer) in the state $R{\left(k\frac{2\pi}{K}+\phi_0\right)}\ket{+_{N,\alpha}}$, for some known offset angle $\phi_0$? In this case, we employ a ``grouped" version of the canonical phase measurement, with measurement operators
\begin{equation}\label{eq:DPmeas}
	\Pi^{\textrm{DP}}_K(k;\phi_0)\equiv \int_{{\left(k-\tfrac{1}{2}\right)}\tfrac{2\pi}{K}+\phi_0}^{{\left(k+\tfrac{1}{2}\right)}\tfrac{2\pi}{K}+\phi_0} \upd\phi~\Pi(\phi),
\end{equation}
$k=0,1,\ldots, K$. We refer to this as the discrete phase (DP) measurement. Implementation-wise, this grouping can be achieved simply by carrying out the original continuous-parameter canonical phase measurement, and then classically grouping the results into the discrete angular intervals.

We also need a measurement to distinguish between the conjugate-basis code states $\ket{\pm_{N,\alpha}}$; we refer to this as the $\overline X$ measurement. Since the $\ket{\pm_{N,\alpha}}$ states are orthogonal, there is a measurement that perfectly distinguishes between them. However, that measurement is built from the $\ket{\pm_{N,\alpha}}$ states and hence code dependent. It is useful to have an $\alpha$-independent $\overline X$ measurement, albeit an approximate one, applicable for different orders $N$. 
As noted in Ref.~\cite{grimsmo2020quantum}, the $\ket{+_{N,\alpha}}$ and $\ket{-_{N,\alpha}}$ states have rather well-defined phases at even and odd, respectively, multiples of $\frac{\pi}{N}$. This suggests the use of the canonical phase measurement as an approximate $\overline X$ measurement: We use a grouped discrete phase measurement to guess if $k$ is even or odd, i.e., the two-element measurement,
\begin{equation}\label{eq:x_meas}
	\Pi_{N,\pm}(\phi_0) \equiv \sum_{k\in\mathbb{Z}_{N,\pm}} \Pi^{\textrm{DP}}_{2N}(k;\phi_0).
\end{equation}
Here, $\phi_0$ can be used to accommodate any systematic rotation offset in the $\ket{\pm_{N,\alpha}}$ states. One can show that the error incurred by using the canonical phase measurement in this manner to distinguish between the $\ket{\pm_N}$ states is exponentially suppressed with $\alpha^2$ (see App.~\ref{app:phase_meas}). This two-element discrete phase measurement hence works as a code-independent (apart from the $N$-dependent classical grouping) $\overline X$ measurement. 

\begin{figure*}
	\centering
	\includegraphics[scale=0.55]{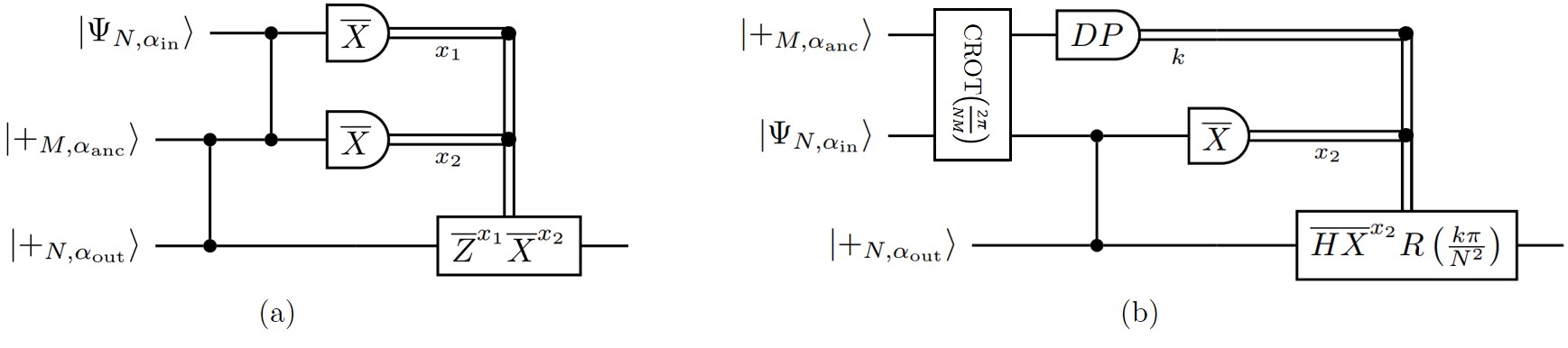}
	\caption{The EC gadgets studied in this work, as introduced in Ref.~\cite{grimsmo2020quantum}. (a) The Knill EC gadget: Each horizontal line represents a bosonic mode, referred to as the input (upper), ancillary (middle), and output (lower) modes. The input and output modes are encoded in an order-$N$ RSB code; the ancillary mode is encoded in an order-$M$ RSB code, with $M$ possibly unequal to $N$. The gates indicated by a vertical line with black dots at its two ends are $\overline\cz$ gates. The input and ancillary modes terminate in $\overline X$ measurements, and classical outcomes $x_1$ and $x_2$ control the recovery on the output mode. (b) The hybrid EC gadget: The three modes are the ancillary (upper), input (middle), and output (lower) modes. A $\crot$ gate is applied on the input and ancillary modes, followed by an $\overline X$ and a discrete phase (DP) measurements on the input and ancillary modes, respectively. A recovery controlled by the classical outcomes $k$ and $x_2$ is finally applied to the output.}
	\label{fig:eccircuits}
\end{figure*}

\subsection{Error correction gadgets}
\label{sec:eccircuits}

Finally, we review the EC gadgets introduced in Ref.~\cite{grimsmo2020quantum}, referred to as the Knill (following Ref.~\cite{knill2005quantum}) and hybrid EC schemes, designed to correct loss and phase errors on the incoming mode carrying possibly noisy computational data. While both schemes work for general RSB codes, we focus on their description for cat codes.

\subsubsection{Knill EC scheme}
\label{sec:knillec}
Figure~\ref{fig:eccircuits}(a) gives the circuit for the Knill EC scheme. The three bosonic modes entering the circuit are labeled as the \emph{input} (upper), \emph{ancillary} (middle), and \emph{output} (lower) modes. The computational data is carried by the input mode initially. In the absence of errors, the circuit implements a logical-qubit teleportation, transferring the information from the input mode to the output mode.
The output mode thus becomes the input mode in subsequent computational and EC cycles. For simplicity, we assume that the input and output modes are both encoded in the same order-$N$ cat code with amplitude $\alpha$. The ancillary mode can be encoded using a cat code with a different order $M$ and a different $\alpha$ value.

In the presence of errors on the input mode, the Knill EC-gadget can correct those errors, provided they are sufficiently weak. Consider the error $R(\theta)\widehat a^k$, for $k=0,1,\ldots,N-1$, on the input mode entering the circuit. From Eq.~\eqref{eq:errorCROT}, we see that, after the $\overline{\cz}$ gate on the input mode, the error on the input remains unchanged but the ancillary mode now acquires an additional $R{\left(-k\frac{\pi}{MN}\right)}$. The $\overline X$ measurement on the ancillary mode now has to distinguish between $\ket{+_{M,\alpha}}$ and $\ket{-_{M,\alpha}}$ in the presence of the added rotation $R{\left(-k\frac{\pi}{MN}\right)}$. This added rotation, as well as any prior information on the likely $k$ values, can be accommodated by adjusting the offset $\phi_0$ in the discrete phase measurement used to implement the $\overline X$ measurement. Provided $k$ is smaller than $N$ (for a correctable error), the outcome of the $\overline X$ measurement is unaffected, i.e., we recover the same measurement value as if there were no error.
Likewise, the result of the $\overline X$ measurement on the input mode can be restored in the presence of the error. In this case, we have the added error of $R(\theta)\widehat a^k$. The $\widehat a^k$ operator moves $\ket{\pm_{N,\alpha}}$ radially in the phase space, without any rotation. For large $\alpha$ and $k$ not too large, this only mildly increases the measurement error probabilities, and we note that the teleportation circuit restores $\alpha$ to the original value in the output mode. For the extra rotation due to $R(\theta)$, the offset $\phi_0$ can again be adjusted to recover the outcome of the $\overline X$ measurement.

Even when $\alpha$ is not too large, and when we have prior information about the noise statistics, a maximum-likelihood decoder can be used to infer the errors and implement the correct recovery operator $\overline Z^{x_1}\overline X^{x_2}$ in the presence of noise \cite{grimsmo2020quantum}. We note that this operator can be implemented virtually by simply recording the measurement outcomes: It is a Pauli operator, and since the EC gadget involves only Clifford gates and Pauli measurements, the effect of the recovery operator is a simple relabeling of the measurement outcomes at the end of the computation, i.e., a Pauli-frame change.

\subsubsection{Hybrid EC scheme}
\label{sec:hybridEC}
The circuit for the hybrid EC scheme, inspired in part by the Steane EC scheme for qubit codes \cite{steane1997active}, is given in Fig.~\ref{fig:eccircuits}(b). As in the Knill EC scheme, the input (middle) and output (lower) modes are encoded with an order-$N$ cat code; the ancillary (upper) mode is encoded with an order-$M$ cat code, with $M$ possibly different from $N$, and possibly a different $\alpha$ value. 
In the absence of errors, $\crot{\left(\frac{2\pi}{MN}\right)}$ acts like the identity on the ancillary-input state. The subsequent phase measurement hence gives $k=0$. The $\overline{\cz}$ gate followed by the $\overline X$ measurement and the correction gate implements a one-bit teleportation that transfers the code state $\ket{\Psi_{N,\alpha}}$ from the input mode to the output mode.

In the presence of error $R(\theta)\widehat a^k$ on the input mode entering the circuit, the $\crot{\left(\frac{2\pi}{MN}\right)}$ gate propagates [see Eq.~\eqref{eq:errorCROT}] a rotation $R{\left(-k\frac{2\pi}{MN}\right)}$ to the ancillary mode. The error on the input mode remains unchanged, and after the $\overline{\cz}$ gate, the output mode acquires also an added rotation $R{\left(-k\frac{\pi}{N^2}\right)}$. Provided $k$ is no larger than $N$, the discrete phase measurement on the ancillary mode can correctly estimate $k$. The $\overline X$ measurement on the input mode works in the same manner as in the Knill EC case, able to correctly distinguish between $\ket{\pm_{N,\alpha}}$ even in the presence of the error. Armed with the value of $k$ and the outcome $x_2$ of the $\overline X$ measurement, the $\overline{HX}^{x_2}$ operator on the output mode completes the teleportation, while the $R{\left(k\frac{\pi}{N^2}\right)}$ rotation fixes the propagated phase error. As in the Knill EC scheme, a maximum-likelihood decoder can be used for better error inference when $\alpha$ is not large, and the final recovery gate on the output can also be implemented virtually.

\section{Formal fault tolerance of the EC gadgets}
\label{sec:ft_analysis}

We first discuss the formal fault tolerance of the EC gadgets proposed in Ref.~\cite{grimsmo2020quantum}, under circuit-level noise. Because of the circuit simplicity and the mild error-propagation properties of the $\crot$ gates, both gadgets are fault tolerant under circuit-level noise, i.e., they function correctly even if there is some (small) noise in the circuit components. Of course, the capacity for correcting input errors is significantly reduced compared to the ideal-circuit situation analyzed in Ref.~\cite{grimsmo2020quantum}; we illustrate this in the next section through numerical simulations. Here, we content ourselves first with the mathematical proof of the fault tolerance of the circuits.

For standard qubit codes, we speak of the correction capacity of a code as its ability to correct errors on up to some number $t$ of physical qubits on which the code is built; we say that the code corrects $t$ errors for short. Fault tolerance properties of circuits constructed for such codes are also defined in terms of number of faults, leading to an associated number of errors, that can be tolerated in the circuit. Specifically, for an EC gadget, our interest here, we say that it is fault-tolerant against $t$ faults if it satisfies the following property, for a code that corrects errors on $t$ or fewer (physical) qubits (see, for example, Ref.~\cite{aliferis2007level}):
\begin{quote}
	\textit{If there are $s$ errors in the input to the EC gadget and $r$ faults in the gadget itself such that $s+r\leq t$, the output of the gadget differs from the correct codeword by at most $r$ errors.}
\end{quote}
Here, the ``correct codeword" refers to the state in the code space obtained from passing the input [with the $s(\leq t)$ errors] through an ideal (i.e., no noise) recovery map that removes all errors. In other words, passing the output of a noisy EC gadget with the above property through an ideal recovery gives the same state as if one had passed the input through the ideal recovery circuit. Faults here refer to malfunctioning operations (including the waiting identity operations), e.g., a gate that did not function as designed, leading to errors on the qubits involved in that operation.
This fault-tolerance property encapsulates two natural desiderata for good EC-gadget design: (i) If the EC gadget is not too faulty, the number of errors in its output should also be small, so that the output contains only correctable errors removable by an ideal recovery; (ii) an EC gadget that is not too faulty should still be able to remove some errors, so that the number of errors in its output is limited only by how faulty it is, not by how many errors were in the input.

Here, we need to extend the fault-tolerance property to the situation of bosonic codes with continuous phase errors. We note the related statement of fault-tolerance requirements in the recent Ref.~\cite{Xu2024FToperation} for cat codes with discrete-variable ancillas; here, our ancillas are themselves bosonic modes and we focus on the EC gadgets at hand, but the underlying principles for fault-tolerant error correction are similar. Specifically, the relevant error model here is one where we have $k$ loss errors and some total angle $\theta$ for the phase error. Following our earlier discussion of the correction capacity of the cat code [see Eq.~\eqref{eq:QEC_E}], an order-$N$ cat code can correct $k\in[0,N)$ loss errors and a phase error $\theta\in(-\pi/N,0]$. Here, we have set $k_0=0$; the range of phase-error angles is also taken to be negative (i.e., $\theta_0=-\pi/N$), to capture the worst-case scenario where the CROT-propagated phase error from losses accumulate in the same direction. The discussion can be generalized to other values of $k_0$ and $\theta_0$. We demand that an EC gadget is considered fault tolerant for such a bosonic-code situation if it satisfies the following property:
\begin{quote}
	\underline{EC-FT property}. \textit{If there are $k_I$ loss errors and a phase error of angle $\theta_I$ in the input to the EC gadget, and $k_G$ loss errors and phase errors of total angle $\theta_G$ in the gadget itself, such that $k_I+k_G<N$ and $|\theta_I+\theta_G+\theta_I'+\theta_G'|<\frac{\pi}{N}$, the output of the gadget differs from the correct codeword by at most $k_G$ loss errors and a phase error of magnitude $\leq|\theta_G+\theta_G'|$. Here, $\theta_I'$ and $\theta_G'$ are, respectively, phase errors at the output that arise from propagating the $k_I$ loss errors in the input and the $k_G$ loss errors in the gadget through the EC gadget}.
\end{quote}
The EC-FT property automatically ensures that the output of the gadget has errors within the capacity of the underlying order-$N$ cat code and are hence correctable. Here, we have assumed that the EC gadgets can propagate loss errors into phase errors, but not vice versa; this is true of the Knill and hybrid EC gadgets.

In App.~\ref{app:ft}, we show that the Knill and hybrid EC gadgets satisfy the EC-FT property, and can hence be considered formally fault tolerant even in the presence of faults in the gadget itself. The proof is straightforward and involves keeping track of how errors accumulate through the circuit elements into the output. 

The proof also gives clues on the optimal choice of code order $M$ for the ancillary mode when requiring fault tolerance of the EC gadgets. For the Knill EC gadget, while $M$ can formally be arbitrary, the best error tolerance is achieved when $M=N$, with $N$ the code order for the input mode; see App.~\ref{app:ft_knill}. For the hybrid EC gadget, choosing $M=1$ turns out to give the best error tolerance, and the amplitude $\alpha$ for that code should be made as high as possible; see App.~\ref{app:ft_hybrid}. The reason is that, for $M=1$, the propagated error from the ancillary mode to the input mode is equivalent to the identity operation on the code space and thus causes no harm, while $M=1$ with a large $\alpha$ maximizes the distinguishability between the $N$ phase states, caused by loss errors in the input. 
We make use of the lessons learned here on the choices of $M$ in our numerical simulations in the next section.

We note that our EC-FT property considers only a limited number of input errors ($k_I$ no larger than $N$). In Ref.~\cite{aliferis2007level} (see also Ref.~\cite{AGP2006}), there are additional conditions for when the input errors are arbitrary (e.g., Property 0 of Ref.~\cite{AGP2006}). Those conditions are needed only in the situation of concatenated codes, to remove the appearance of leakage outside of the qubit space problematic for the next level of concatenation. In our current context, we do not have code concatenation, and hence a condition only for limited input errors suffices.

\section{Performance of the EC gadgets: Numerical simulations}
\label{sec:num_sim}

As with any fault-tolerance analysis, the true test of the efficacy of the circuit design lies in its performance in the presence of noise, and the resulting threshold noise level below which the EC circuit achieves a net removal of noise despite the imperfections in its components. Here, we perform numerical simulations of the EC gadgets for the memory task, i.e., storage of quantum information without the application of any nontrivial computational gates. Error correction, implemented by an EC gadget, is periodically done to attempt to remove errors. Below, we describe first the simulation setup and noise model, and then numerically examine the efficacy of the error correction in preserving the stored information, under different settings.

\subsection{Simulation setup}
\label{sec:setup}

\newcommand{\Twait}{\tau_\mathrm{wait}}
\newcommand{\gammaL}{\gamma_\mathrm{loss}}
\newcommand{\gammaD}{\gamma_\mathrm{ph}}
\newcommand{\inF}{\mathrm{in}\cF}
\newcommand{\inFbm}{\mathrm{in}\cF_\mathrm{bm}}

\begin{figure}[h]
	\centering
	\includegraphics[width=\columnwidth]{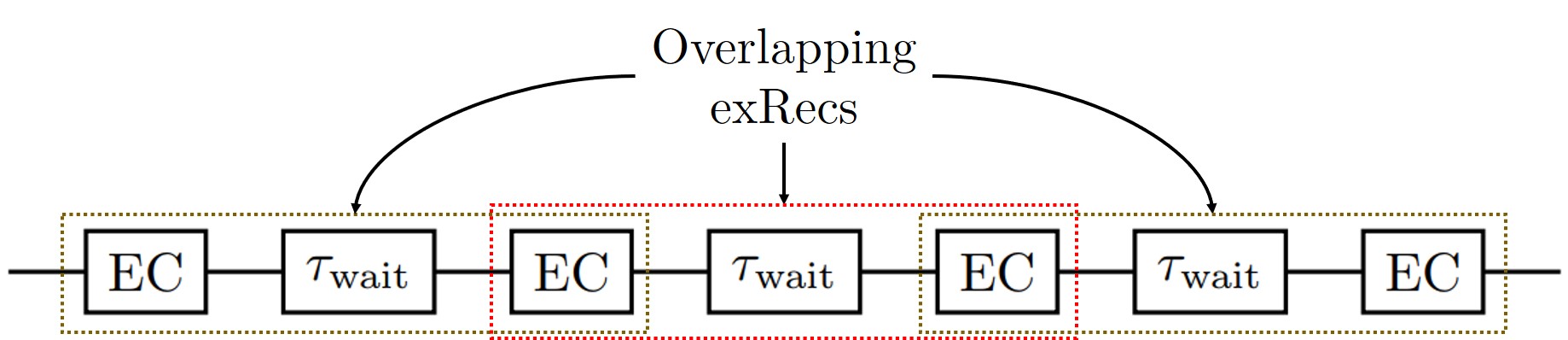}
	\caption{The error-corrected memory circuit as a sequence of overlapping exRecs. Each exRec comprises two EC gadgets separated by a waiting period $\Twait$.}
	\label{fig:extended_gadget}
\end{figure}

To understand the effectiveness of periodic error correction, we only need to analyze the behavior of a modular unit, namely the \emph{extended memory rectangle}, or just ``exRec" for short (following terminology introduced in Ref.~\cite{aliferis2007level}); see Fig.~\ref{fig:extended_gadget}. The exRec comprises two---the leading (earlier) and trailing (later)---EC gadgets separated by a waiting period $\Twait$. Although originally developed for qubit codes, the exRec approach \cite{aliferis2007level} to fault-tolerance proofs applies to our current RSB-codes situation, if the EC gadgets satisfy the EC-FT property. Specifically, we break up the entire memory circuit, comprising a string of EC gadgets separated by waiting periods, into consecutive overlapping exRecs; see Fig.~\ref{fig:extended_gadget}. Desideratum (ii) above, underlying the formulation of the EC-FT property, ensures we need look no further back than the leading EC to capture the errors entering the trailing EC, giving the exRec as the modular unit of consideration. The correctness of the entire circuit is ensured by the correctness of each exRec, and the failure probability of the full memory circuit is then given by the probability that at least one of the exRec fails (we define below what it means for an exRec to fail). This reduces the problem to examining the failure probability of a single exRec.

Within a single exRec, faults are inserted randomly at different locations according to the noise model described below (see Sec.~\ref{sec:noise}). If there are no faults in the trailing EC, that EC-gadget functions as an ideal EC and the output state from the exRec lies in the code space. This output state is the correct logical state (i.e., equal to the no-error input state) if the faults inserted prior to the trailing EC lead to correctable errors removed by the trailing EC; otherwise, a logical error may occur. In the presence of faults in the trailing EC, that EC-gadget can function incorrectly, and the output state may no longer be in the code space. There are then two cases: Case (i)-- The output differs from the no-error input state by a correctable error. That error can potentially be removed by a subsequent EC gadget. We hence do not consider this a failure of the ex-Rec. Case (ii)-- The output differs from the no-error input by an uncorrectable error. If so, even a further ideal EC gadget cannot remove the error and the input state has suffered irreparable damage. This is then a failure of the ex-Rec. 
For convenience in separating the two cases, in our simulations, we pass the output of the faulty exRec through an additional (ideal) mathematical recovery map (see App.~\ref{app:Sim1}) that maps Case (i) to the no-error state, and Case (ii) to a state with a logical error. Below, the exRec is understood to include this additional mathematical recovery, and the output is always in the code space.

We repeat the simulation many times, each time with a random instance of fault insertions, and we compute the ratio $R$ of the average infidelity (Haar-averaged over pure input states; see App.~\ref{app:Sim2}) to a benchmark infidelity. In the average infidelity, we are comparing the output of the exRec with the no-error input state. For the benchmark, we make the standard choice in bosonic codes literature: The information is carried by the subspace spanned by the $\ket 0$ and $\ket 1$ Fock states. In the benchmark situation, we consider storage of the information for the waiting period $\Twait$, during which loss and phase errors can occur. The break-even point is the operating point where $R=1$, i.e., the infidelities are equal with and without error correction;  when $R<1$, doing error correction yields a lower logical error.

In our analysis, we optimize $R$ over the adjustable parameters characterizing the EC gadgets, namely, the amplitudes $\alpha_\textrm{in}$ and $\alpha_\textrm{anc}$ for the input/output and ancillary modes, and the offsets $\phi_{0;\textrm{in}}$ and $\phi_{0;\textrm{anc}}$ for the discrete phase or $\overline X$ measurements on the input and ancillary modes. For experimental relevance, we limit the $\alpha$ values to the range $\alpha_\textrm{in},\alpha_\textrm{anc}\in[0,9]$. In the same EC gadget, $\alpha_\text{in}$ and $\alpha_\text{anc}$ can take on different values, but those values are kept constant for the two EC gadgets in the exRec.

\subsection{Circuit-level noise model}	
\label{sec:noise}

Earlier, when exploring the fault tolerance of the EC gadgets, we discussed the loss and phase errors in terms of the number $k$ of loss errors, and the angle $\theta$ for a phase error. In our simulations, we need a concrete noise model that specifies how often these errors occur. As is standard in the bosonic-codes literature (see, for example, Refs.~\cite{grimsmo2020quantum} and \cite{joshi2021quantum}), we assume loss errors arise from a loss channel, described as the completely positive and trace-preserving map,
\begin{align}\label{eq:loss_kraus}
	\cA_{\gammaL}(\,\cdot\,)&\equiv \sum_{k=0}^\infty A_k(\,\cdot\,)A_k^\dagger\\
	\textrm{with}\quad A_k &\equiv \frac{(1-\upe^{-\gammaL})^{k/2}}{\sqrt{k!}} \upe^{-\frac{1}{2}\gammaL \widehat{n}}\widehat{a}^k.\nonumber
\end{align}
Here, the strength of the loss channel is quantified by the parameter $\gammaL \equiv \kappa t$, for the loss rate $\kappa$ and some time $t$. Phase errors arise from a dephasing channel,
\begin{equation}\label{eq:dephasing_gauss}
	\cD_{\gammaD}(\,\cdot\,) \equiv \int_{-\infty}^{\infty} \upd\theta~p_{\gammaD}(\theta) \upe^{\upi\theta\widehat{n}} (\,\cdot\,) \upe^{-\upi\theta\widehat{n}}.
\end{equation}
Here, the dephasing strength is $\gammaD \equiv \kappa_\textrm{ph} t$, with dephasing rate $\kappa_\textrm{ph}$ and some time $t$. $p_{\gammaD}(\theta)$ specifies the distribution of the phase errors, which we set to be Gaussian in nature: $ p_{\gammaD}(\theta)\equiv  \frac{1}{\sqrt{2\pi\gammaD}}\upe^{-\theta^2/(2\gammaD)}$. 
The overall noise is a simple composition of the two maps: $\cN_{\gammaL, \gammaD}(\cdot) = \cD_{\gammaD} \circ \cA_{\gammaL} (\cdot)$.

For circuit-level noise, we divide the exRec into multiple locations, each location containing a single operation---preparation, gate (including waiting), or measurement. The waiting period $\Twait$ for each mode is each regarded as a single location of duration $\Twait$. The Knill and hybrid EC gadgets contain four types of locations: the preparation of a logical state, a $\textsc{CROT}$ gate, an $\overline{X}$ measurement, and a waiting (or idling) location. Each location in the EC gadget is assumed, for simplicity, to take the same duration $\tau$. Every location in the exRec is subjected to the noise $\cN_{\gammaL,\gammaD}$ with $\gammaL= \gammaL^\textrm{wait}\equiv \kappa\Twait$ and $\gammaD=\gammaD^\textrm{wait}\equiv\kappa_\textrm{ph}\Twait$ for the waiting period, and $\gammaL=\gammaL^\textrm{op}\equiv\kappa\tau$ and $\gammaD=\gammaD^\textrm{op}\equiv\kappa\tau$ for locations in the EC gadgets. In addition, we assume an ordering of the noise and the ideal operations: the noisy preparation of a logical state is modeled as the ideal preparation followed by $\cN_{\gammaL, \gammaD}(\cdot)$; the noisy two-mode $\textsc{CROT}$ gate is modeled as the ideal $\textsc{CROT}$ followed by $\cN_{\gammaL, \gammaD}(\cdot)$ acting independently on each of the two modes; lastly, a noisy measurement is modeled as an ideal measurement preceded by $\cN_{\gammaL, \gammaD}(\cdot)$.

\subsection{Simulation results}
Here, we investigate the performance of the Knill and hybrid EC gadgets numerically under three scenarios: a fixed waiting time $\Twait$ (Sec.~\ref{sec:fix_wait}), an optimized $\Twait$ (Sec.~\ref{sec:vary_wait}), and squeezed cat codes (Sec.~\ref{sec:squeeze}).

\subsubsection{Fixed waiting period}
\label{sec:fix_wait}

\begin{figure*}
	\centering
	\includegraphics[width=0.85\textwidth]{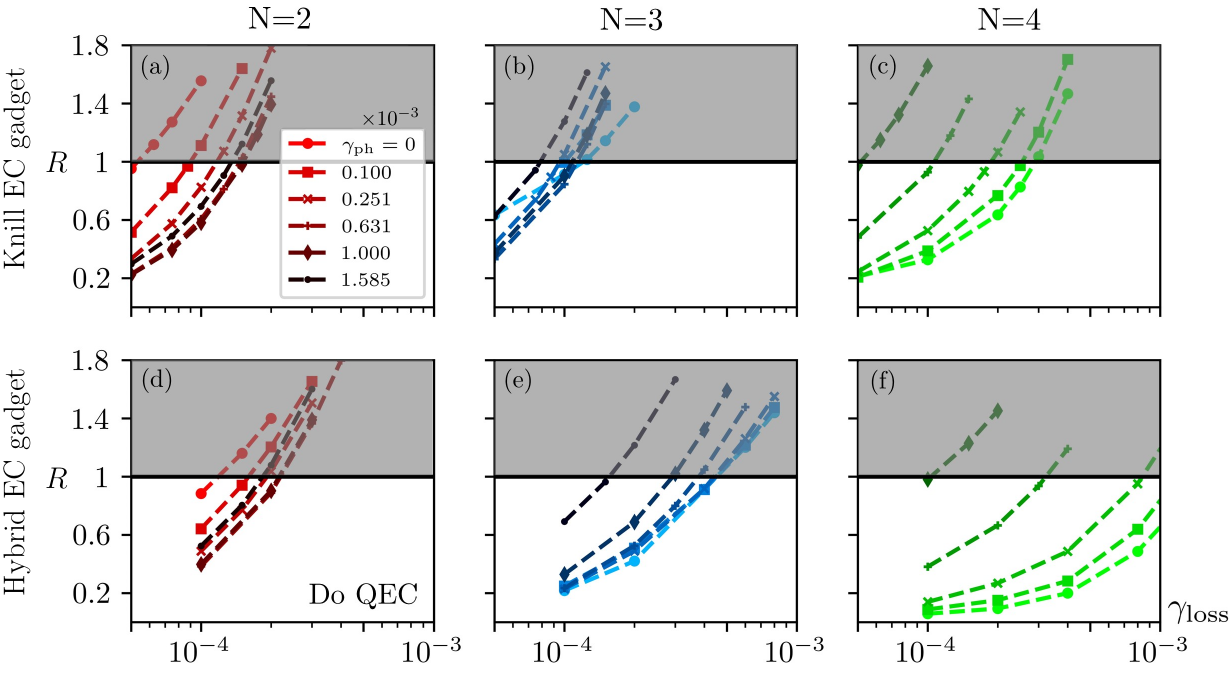}
	\caption{Fixed waiting period: ratio $R$ of the average infidelities of the error-corrected and benchmark situations for the Knill and hybrid EC gadgets. We plot $R$ against loss error probability $\gammaL$, with separate lines of increasingly dark hues for $\gammaD=0,1\times10^{-4}, 2.51\times 10^{-4}, 6.31\times 10^{-4}, 1\times 10^{-3}$, and $1.585\times 10^{-3}$. The black horizontal line gives the $R=1$ break-even point. The white region, where $R<1$, corresponds to where error correction outperforms the benchmark unencoded situation.}
	\label{fig:fixed_wait}
\end{figure*}

\begin{figure}[h!]
	\centering
	\includegraphics[width=\columnwidth]{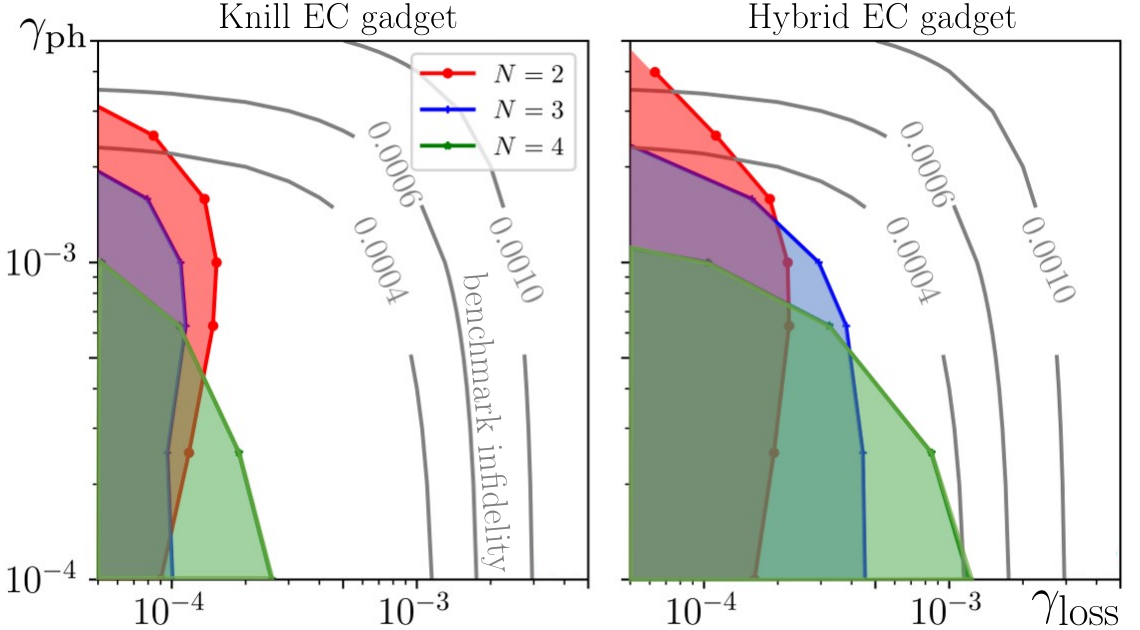}
	\caption{\label{fig:fixed_wait_threshold}Fixed waiting period: regions where error correction outperforms the benchmark unencoded situation. The boundary curves of the shaded regions are the break-even points where $R=1$; the shaded regions give the loss ($\gammaL$) and phase ($\gammaD$) error probabilities where it is better to do error correction. The lines in gray are contours for the benchmark infidelity values.}	
\end{figure}

We begin with the simplest case where the waiting period in between two consecutive EC gadgets is kept equal to the operation time, i.e., we set $\Twait=\tau$, so that $\gammaL^\textrm{op}=\gammaL^\textrm{wait}=\kappa\tau\equiv \gammaL$ and $\gammaD^\textrm{op}=\gammaD^\textrm{wait}=\kappa_\textrm{ph}\tau\equiv \gammaD$. For each pair of noise parameters $(\gammaD,\gammaL)$, we optimize the infidelity ratio $R$ over $\alpha_\textrm{in}$, $\alpha_\textrm{anc}$, $\phi_{0;\textrm{in}}$, and $\phi_{0,\textrm{anc}}$. Figure~\ref{fig:fixed_wait} shows the optimized $R$ for the Knill and hybrid EC gadgets, as a function of $\gammaL$, for different values of $\gammaD$ and different code order $N$, focusing on the region of noise parameters around the break-even points (i.e., where $R$ is close to 1). From Fig.~\ref{fig:fixed_wait}, we can deduce the break-even points; Fig.~\ref{fig:fixed_wait_threshold} shows the regions in the $(\gammaL, \gammaD)$ plane for different $N$ values, where error correction outperforms the benchmark unencoded situation.

From Figs.~\ref{fig:fixed_wait} and \ref{fig:fixed_wait_threshold}, it is clear that the hybrid EC gadget performs significantly better than the Knill EC gadget: For a fixed $N$, and the same values of $(\gammaL,\gammaD)$, the hybrid EC gadget gives a smaller $R$ than the Knill one. We believe this is due to the fact that the hybrid scheme suffers less from noise in the ancillary mode. To understand that, we first note that increasing $\alpha_\textrm{anc}$ reduces the measurement error (see App.~\ref{app:phase_meas}) at the cost of a larger loss probability for the larger mean photon number. For the hybrid EC gadget, this larger loss probability does not affect its overall performance as losses in the ancillary mode propagate to phase errors equivalent to the identity operation on the input mode (due to the choice of $M=1$; see App.~\ref{app:ft_hybrid}). This is different in the Knill case: There, $N$ losses can propagate a $\overline Z$ error to the input and hence, increasing $\alpha_\mathrm{anc}$ can increase the logical error probability. We note that this difference in performance between the two EC gadgets, due to the errors in the ancillary mode, is in sharp contrast to the conclusion of Ref.~\cite{grimsmo2020quantum}. There, noise affects only the input mode, and in that case, the two schemes were shown to perform comparably.

Three further observations are visible in the figures. First, we note the persistence of the code-capacity behavior: A larger $N$ gives increased tolerance to loss errors but decreased tolerance to phase errors. We had recognized this earlier as a property of the cat code itself; here, we see a similar behavior even when the EC gadgets are noisy. This is particularly clear from Fig.~\ref{fig:fixed_wait_threshold}, where the region where the error-corrected case performs better than the benchmark includes, for smaller $N$, larger values for $\gammaD$ and smaller values for $\gammaL$, compared with larger $N$.

Second, for $N=2$, we observe the curious feature that, in the region of small $\gammaD$, increasing $\gammaD$ actually improves the ratio of phase to loss tolerance. However, beyond a certain $\gammaD$, a further increase in $\gammaD$ no longer improves the performance but starts to worsen it. The turning point occurs at $\gammaD\sim 0.8-1\times10^{-3}$ for both EC gadgets. This is visible in Figs.~\ref{fig:fixed_wait}(a) and (d), where the lines first move rightwards as $\gammaD$ increases, only to start moving left again as $\gammaD$ continues to grow; this gives the corresponding feature in Fig.~\ref{fig:fixed_wait_threshold} where the border of the red region bends inwards for values of $\gammaD$ smaller than around $1\times 10^{-3}$. This behavior can be explained by considering the properties of the $N=2$ cat code: The code has high tolerance to phase errors, allowing it to perform well in the low $\gammaD$ region. As $\gammaD$ increases, the error-corrected fidelity is not significantly affected. However, the fidelity for the no-EC benchmark case is significantly reduced, leading to an overall decrease in the ratio $R$. For $N>2$, we do not observe the same behavior; instead, increasing $\gammaD$ consistently worsens the performance of the EC schemes, at least for the region of noise parameters considered here.

Third, from Fig.~\ref{fig:fixed_wait_threshold}, we see that the $N=3$ cat code achieves break even at $(\gammaL, \gammaD) \sim (10^{-4},  10^{-3})$ [optimized amplitudes: $(\alpha_{\mathrm{in}}, \alpha_{\mathrm{anc}}) = (7.5, 6.6)$]  for the Knill scheme and at $(\gammaL, \gammaD) \sim (5\times10^{-4},  10^{-3})$ [optimized amplitudes: $(\alpha_{\mathrm{in}}, \alpha_{\mathrm{anc}}) = (6.3, 7.8)$] for the hybrid scheme. These numbers are significantly lower than the corresponding estimates in the code-capacity \cite{grimsmo2020quantum} or measurement-noise-only \cite{hillmann2022performance} models, where the break-even points are at $\gammaL = \gammaD \sim 10^{-2}$. This emphasizes the importance of considering the performance of the EC gadgets in the presence of the full circuit-level noise used here, and suggests that errors arising from the ancillary modes can significantly affect the circuit performance. In the following two sections, we explore how we can relax the stringent requirements for break even by first optimizing the waiting time, and then by using squeezed cat codes.

\subsubsection{Optimized waiting period}
\label{sec:vary_wait}

For a memory task with noisy EC gadgets, we can find the optimal rate at which error correction should be performed to attempt to remove errors in the memory. Doing error correction too frequently can add, rather than remove, noise due to the errors in the EC gadgets themselves; infrequent error correction gives an increased possibility of uncorrected errors accumulating into irreparable damage on the stored information. We should thus vary $\Twait$ (rather than fixing it at $\tau$ as in the previous section) to minimize the ratio $R$. In our simulations, this amounts to assigning different $\gammaL$ and $\gammaD$ values for the waiting period and for locations in the EC gadgets. Specifically, we write $\gammaL^\textrm{wait}\equiv\kappa\Twait$ and $\gammaD^\textrm{wait}\equiv\kappa_\textrm{ph}\Twait$ for the waiting period; those for locations in the EC gadgets are as before, i.e., $\gammaL^\textrm{op}\equiv\kappa\tau\equiv\gammaL$ and $\gammaD^\textrm{op}\equiv\kappa_\textrm{ph}\tau\equiv \gammaD$.

\begin{figure}
	\centering
	\includegraphics[width=0.49\textwidth]{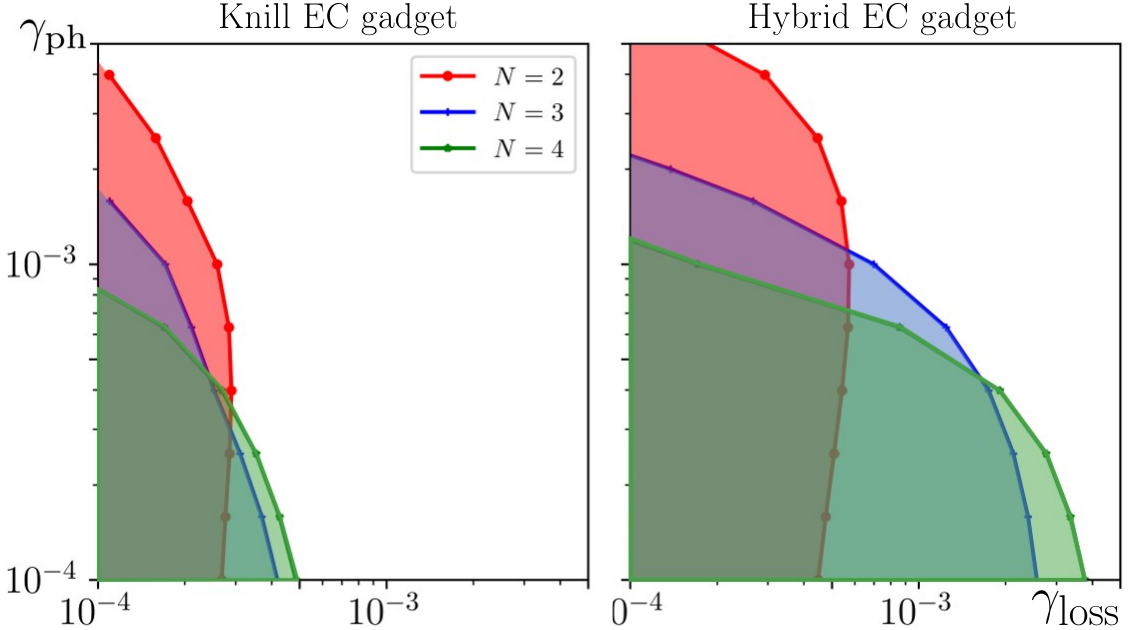}
	\caption{\label{fig:varied_wait_threshold}Optimized waiting period: regions where error correction outperforms the benchmark unencoded situation, plotted for the error probabilities $\gammaL$ and $\gammaD$ for the EC operations. The boundary curves of the shaded regions are the break-even points where $R=1$; the shaded regions give the noise parameters where it is better to do error correction.}	
\end{figure}

Figure~\ref{fig:varied_wait_threshold} gives the below-break-even regions for the two EC gadgets. In this varying-$\Twait$ scenario, we can draw observations very similar to those discussed in the fixed-waiting-time scenario. However, both EC gadgets clearly achieve break-even at larger noise values, compared to the fixed waiting time situation. Specifically, for a given phase error probability $\gammaD^\textrm{op}$, the break-even point for $\gammaL^\textrm{op}$ is improved by a factor of 2--4 for the Knill EC gadget, and by 3--5 for the hybrid EC gadget. In fact, with the loss probability of $\gammaL=10^{-3}$, the hybrid scheme can achieve break even at a phase error probability as high as $\gamma_{\phi} \approx 7\times 10^{-4}$, values attainable in current hardware \cite{sivak2023real}. The corresponding optimal amplitude for the input mode is also quite reasonable, at $\alpha_{\text{in}} \approx 4$, equivalent to a mean photon number $\overline{n}\approx 16$. We note that the optimal $\Twait$ varies from case to case, but the values range from a few to a few-tens times $\tau$.

\subsubsection{Squeezed cat codes}
\label{sec:squeeze}
Finally, we explore the possibility of adding squeezing to the code states to further enhance the performance of the EC gadgets. For the squeezed cat code, a squeezed coherent state is used as the primitive state, in place of a coherent state as in the regular cat code. A squeezed coherent state is defined as
\begin{equation}\label{eq:sqz_state}
	\ket{\alpha, r, \varphi} \equiv \widehat{D}(\alpha)\widehat{S}(r,\varphi)\ket{\textrm{vac}},
\end{equation}
where $\widehat{D}(\alpha) = \upe^{(\alpha\widehat{a}^{\dagger} - \alpha^{*} \widehat{a})}$ is the displacement operator, $\widehat{S}(r, \varphi) = \upe^{\frac{1}{2}(z(\widehat{a}^{\dagger})^2 - z^{*}\widehat{a}^2)}$ is the squeezing operator with $z\equiv r\upe^{2\upi\varphi}$, and $\ket{\textrm{vac}}$ is the vacuum state. The angle $\varphi$ determines the quadrature $\Delta X_\varphi^2$ with minimal variance. The squeezing parameter $r$ determines the amount of squeezing, with the squeezing factor (in dB) defined as
\begin{equation}
	f \equiv -10 \log_{10}\left(\frac{\Delta \widehat{X}_{\varphi} ^2}{\Delta \widehat{X}_{\text{vac}}^2}\right) = -10 \log_{10} \upe^{-2r} \approx 8.68r,
\end{equation}
where $\Delta \widehat X_\varphi^2$ and $\Delta X_\mathrm{vac}^2$ are, respectively, the quadrature variances of the squeezed and vacuum states. 
Under a phase error, a squeezed coherent state is rotated about the origin, 
\begin{equation}\label{eq:sqz_rotation}
	\widehat{R}(\theta) \ket{\alpha, r, \varphi} = \ket{\alpha \upe^{\upi\theta}, r, \varphi + \theta}
\end{equation}
Under a loss error, a squeezed coherent state becomes
\begin{align}\label{eq:sqz_loss}
	\begin{split}
		\widehat{a} \ket{\alpha, r, \varphi} = \hspace{1mm} &\alpha \ket{\alpha, r, \varphi} - \\
		&\upe^{2\upi\varphi}\sinh(r) \widehat{D}(\alpha)\widehat{S}(r,\varphi) \widehat{a}^\dagger \ket{\textrm{vac}}.
	\end{split}
\end{align}

The potential of incorporating squeezing to enhance protection against loss and phase errors has been explored in several theoretical studies \cite{schlegel2022quantum, teh2020overcoming, hillmann2023quantum, xu2023autonomous}, with experimental demonstrations for the order-$1$ cat code \cite{le2018slowing, pan2023protecting}. In these schemes, when dealing with high losses and small amplitudes, squeezing along the radial direction (i.e., setting $\varphi = 0$ in Eq.~\ref{eq:sqz_state}) mitigates the effects of loss and extends the lifetime of a cat state. 

The Knill and hybrid EC gadgets, applicable for arbitrary RSB codes, work also for the squeezed cat code, and can be analyzed in a manner similar to the earlier regular cat code, with the main difference arising from how errors affect the squeezed code states [see Eqs.~\eqref{eq:sqz_rotation} and \ref{eq:sqz_loss}]. Since measurement errors, stemming from constraints on the achievable amplitudes, is one limiting factor for the performance of these gadgets, we find that employing squeezing along the phase direction---specifically, choosing $\varphi = \frac{\pi}{2}$ in Eq.~\ref{eq:sqz_state}---offers improved performance by reducing the measurement error rate. We thus set $\varphi = \frac{\pi}{2}$ in our simulations and optimize the squeezing parameter in the practically relevant range of $r\in[0,1.5]$, equivalent to the range $f\in[0,13]$dB. 

\begin{figure}
	\centering
	\includegraphics[width=\columnwidth]{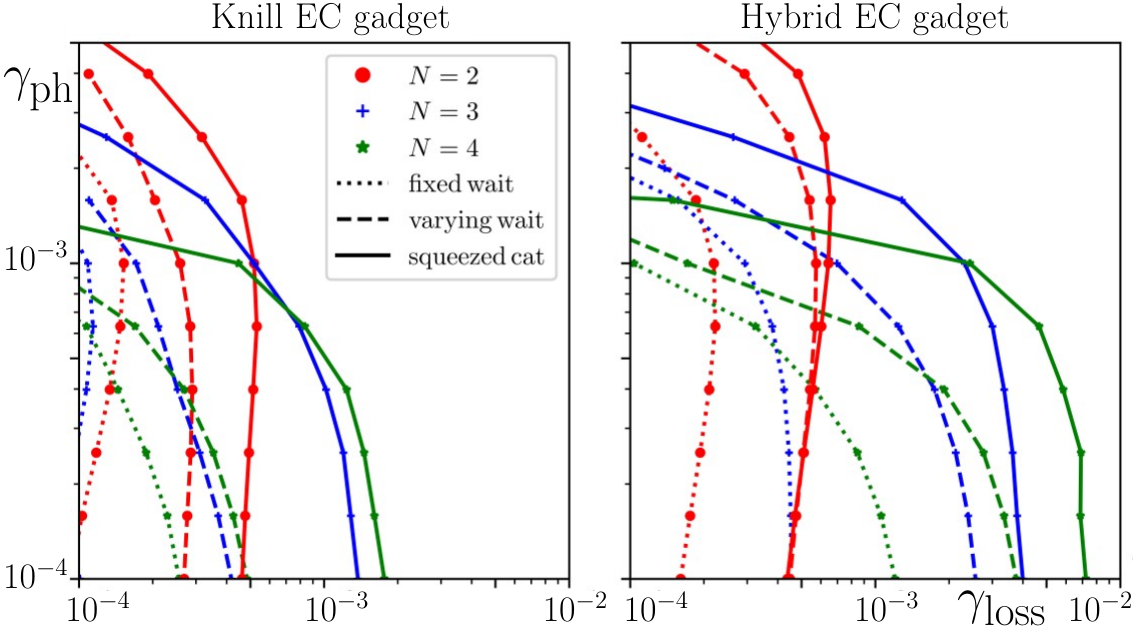}
	\caption{\label{fig:squeeze_threshold} Squeezed cat codes with optimized waiting period: regions where error correction outperforms the benchmark unencoded situation, The curves give the break-even points where $R=1$; the data for the earlier regular cat codes situations are shown for comparison. The region to the left and below each curve is where it is better to do error correction ($R<1$).}
\end{figure}

Figure \ref{fig:squeeze_threshold} shows the break-even points for the Knill and hybrid EC gadgets for the squeezed cat code, with optimized waiting time; our earlier results for the regular cat code are plotted together for comparison. Notably, the squeezed cat code extends the break-even boundaries for both loss and phase errors. For a given $\gammaD$ value, the improvement in the loss threshold for order-$2$ code is about $3$ times that for the fixed-waiting-time and regular cat code scenario. Meanwhile, the improvement for orders-$3$ and $4$ can be as high as an order of magnitude.   
As higher-order RSB codes have smaller tolerance for phase errors, they are more susceptible to errors in the canonical phase measurement. Squeezing along the phase direction is expected to yield a larger effect for higher-order codes. We indeed see this enhancement in Fig.~\ref{fig:squeeze_threshold}, where the threshold improvement is more pronounced for $N=3$ and $N=4$, compared to $N=2$.

Finally, we note that the optimized parameters obtained in our analysis indicate that we do not need extremely high squeezing factors to see the benefits of squeezing. For both EC gadgets, the optimal squeezing factor is typically less than $10$ dB. For example, for $\gammaL=1.6\times10^{-3}$ and $\gammaD=10^{-3}$, the hybrid scheme can achieve $R=0.78$ for an $N=3$ squeezed cat code with an amplitude of $\alpha\approx 4$ and a squeezing factor of $f\approx 6$dB. At these noise levels, the regular cat code is above the threshold, with $R=2.12$. These parameter values are within the reach of current superconducting hardware \cite{pan2023protecting}.

\section{Discussion}
\label{sec:concl}

In this work, we have carefully defined the notion of fault tolerance in the context of RSB codes, and further analyzed the performance of the EC gadgets proposed in Ref.~\cite{grimsmo2020quantum} under circuit-level noise.
Despite the fact that Ref.~\cite{grimsmo2020quantum} considered only input errors, our analysis here demonstrates the fault tolerance of their judicious designs even under circuit-level noise. The tolerance to errors is naturally not as high, with a deterioration of one to two orders of magnitude in the break-even values, but we show how to relax the noise requirements through wait-time optimization and the use of squeezing in the cat codes. That the hybrid EC gadget performs better than the Knill one, as well as the choices of code-order in the ancillary mode for maximal noise robustness, are aspects visible only in our circuit-level analysis. 

One source of noise not considered in our work is the possibility of control imperfections in the two-mode CROT gate that can lead to correlated errors in both participating modes. In our analysis here, we considered only independent errors on the two modes, even when they are coupled by the same CROT gate. This is natural when considering the usual source of background noise on the bosonic modes, as we do here, but may not reflect control noise that can cause simultaneous errors on both modes. A better knowledge of how the CROT gate is implemented physically, and hence the likely nature of control errors, can allow for such an extended analysis in future work.

Another source of imperfections not accounted for in our current model is state-preparation noise that can scale with the code-order $N$. Here, we assume that the preparation of any code state, at the start of the EC gadget, is subjected to the same noise, regardless of $N$. This reflects the current typical experimental procedure for preparing arbitrary bosonic states via optimal-control pulses with a fixed time. In practice, it may be that the larger $N$ states are harder to prepare with high fidelity, requiring more complicated pulse shaping and hence possibly larger imperfections, or that the achievable fidelity for the given fixed pulse length is lower. Preliminary investigations into preparation noise that scales with $N$ suggest that the impact on the break-even values can be significant, but further work is necessary for concrete conclusions if this does emerge as a critical experimental constraint.

Our analysis here centres around the performance of the EC gadgets and hence on the storage task only. We expect a similar approach to apply in the analysis of logical operations on bosonic codes, to complete the circuit-level fault tolerance analysis of RSB codes for arbitrary quantum computation. In particular, one might be able to adapt the analysis of the recent work of Matsuura et al.~\cite{Matsuura2024CVFT}, which showed the fault tolerance of GKP codes under general noise using qubit fault-tolerance proofs, to show fault tolerance in the RSB situation for general computational tasks.

\acknowledgements
We acknowledge helpful discussions with Yvonne Gao, Amon Kasper, and Adrian Copetudo. This work was supported by the Singapore Ministry of Education (through Grant No.~MOE-T2EP50222-0004), and by the National Research Foundation, Singapore and A*STAR under its CQT Bridging Grant.

\bibliographystyle{quantum}
\bibliography{FT-RSB.bib}

\begin{thebibliography}{10}

\bibitem{shor1996fault}
Peter~W Shor.
\newblock ``Fault-tolerant quantum computation''.
\newblock In Proceedings of 37th conference on foundations of computer science.
\newblock Pages 56--65.
\newblock IEEE~(1996).
\newblock  url:~\url{https://doi.org/10.1109/SFCS.1996.548464}.

\bibitem{steane1997active}
Andrew~M Steane.
\newblock ``Active stabilization, quantum computation, and quantum state
  synthesis''.
\newblock Physical Review Letters {\bf 78}, 2252~(1997).
\newblock  url:~\url{https://doi.org/10.1103/PhysRevLett.78.2252}.

\bibitem{gottesman1998theory}
Daniel Gottesman.
\newblock ``Theory of fault-tolerant quantum computation''.
\newblock Physical Review A {\bf 57}, 127~(1998).
\newblock  url:~\url{https://doi.org/10.1103/PhysRevA.57.127}.

\bibitem{knill2005quantum}
Emanuel Knill.
\newblock ``Quantum computing with realistically noisy devices''.
\newblock Nature {\bf 434}, 39--44~(2005).
\newblock  url:~\url{https://doi.org/10.1038/nature03350}.

\bibitem{fowler2012surface}
Austin~G Fowler, Matteo Mariantoni, John~M Martinis, and Andrew~N Cleland.
\newblock ``Surface codes: Towards practical large-scale quantum computation''.
\newblock Physical Review A {\bf 86}, 032324~(2012).
\newblock  url:~\url{https://doi.org/10.1103/PhysRevA.86.032324}.

\bibitem{egan2021fault}
Laird Egan, Dripto~M Debroy, Crystal Noel, Andrew Risinger, Daiwei Zhu,
  Debopriyo Biswas, Michael Newman, Muyuan Li, Kenneth~R Brown, Marko Cetina,
  et~al.
\newblock ``Fault-tolerant control of an error-corrected qubit''.
\newblock Nature {\bf 598}, 281--286~(2021).
\newblock  url:~\url{https://doi.org/10.1038/s41586-021-03928-y}.

\bibitem{google2021exponential}
Google~Quantum AI.
\newblock ``Exponential suppression of bit or phase errors with cyclic error
  correction''.
\newblock Nature {\bf 595}, 383--387~(2021).
\newblock  url:~\url{https://doi.org/10.1038/s41586-021-03588-y}.

\bibitem{krinner2022realizing}
Sebastian Krinner, Nathan Lacroix, Ants Remm, Agustin Di~Paolo, Elie Genois,
  Catherine Leroux, Christoph Hellings, Stefania Lazar, Francois Swiadek,
  Johannes Herrmann, et~al.
\newblock ``Realizing repeated quantum error correction in a distance-three
  surface code''.
\newblock Nature {\bf 605}, 669--674~(2022).
\newblock  url:~\url{https://doi.org/10.1038/s41586-022-04566-8}.

\bibitem{zhao2022realization}
Youwei Zhao, Yangsen Ye, He-Liang Huang, Yiming Zhang, Dachao Wu, Huijie Guan,
  Qingling Zhu, Zuolin Wei, Tan He, Sirui Cao, et~al.
\newblock ``Realization of an error-correcting surface code with
  superconducting qubits''.
\newblock Physical Review Letters {\bf 129}, 030501~(2022).
\newblock  url:~\url{https://doi.org/10.1103/PhysRevLett.129.030501}.

\bibitem{google2023suppressing}
Google~Quantum AI.
\newblock ``Suppressing quantum errors by scaling a surface code logical
  qubit''.
\newblock Nature {\bf 614}, 676--681~(2023).
\newblock  url:~\url{https://doi.org/10.1038/s41586-022-05434-1}.

\bibitem{bluvstein2024logical}
Dolev Bluvstein, Simon~J Evered, Alexandra~A Geim, Sophie~H Li, Hengyun Zhou,
  Tom Manovitz, Sepehr Ebadi, Madelyn Cain, Marcin Kalinowski, Dominik
  Hangleiter, et~al.
\newblock ``Logical quantum processor based on reconfigurable atom arrays''.
\newblock Nature {\bf 626}, 58--65~(2024).
\newblock  url:~\url{https://doi.org/10.1038/s41586-023-06927-3}.

\bibitem{chamberland2017overhead}
Christopher Chamberland, Tomas Jochym-O'Connor, and Raymond Laflamme.
\newblock ``Overhead analysis of universal concatenated quantum codes''.
\newblock Physical Review A {\bf 95}, 022313~(2017).
\newblock  url:~\url{https://doi.org/10.1103/PhysRevA.95.029904}.

\bibitem{kim2022fault}
Isaac~H Kim, Ye-Hua Liu, Sam Pallister, William Pol, Sam Roberts, and Eunseok
  Lee.
\newblock ``Fault-tolerant resource estimate for quantum chemical simulations:
  Case study on li-ion battery electrolyte molecules''.
\newblock Physical Review Research {\bf 4}, 023019~(2022).
\newblock  url:~\url{https://doi.org/10.1103/PhysRevResearch.4.023019}.

\bibitem{beverland2022assessing}
Michael~E Beverland, Prakash Murali, Matthias Troyer, Krysta~M Svore, Torsten
  Hoeffler, Vadym Kliuchnikov, Guang~Hao Low, Mathias Soeken, Aarthi Sundaram,
  and Alexander Vaschillo.
\newblock ``Assessing requirements to scale to practical quantum
  advantage''~(2022).
\newblock  url:~\url{https://doi.org/10.48550/arXiv.2211.07629}.

\bibitem{chuang1997bosonic}
Isaac~L Chuang, Debbie~W Leung, and Yoshihisa Yamamoto.
\newblock ``Bosonic quantum codes for amplitude damping''.
\newblock Physical Review A {\bf 56}, 1114~(1997).
\newblock  url:~\url{https://doi.org/10.1103/PhysRevA.56.1114}.

\bibitem{cochrane1999macroscopically}
Paul~T Cochrane, Gerard~J Milburn, and William~J Munro.
\newblock ``Macroscopically distinct quantum-superposition states as a bosonic
  code for amplitude damping''.
\newblock Physical Review A {\bf 59}, 2631~(1999).
\newblock  url:~\url{https://doi.org/10.1103/PhysRevA.59.2631}.

\bibitem{gottesman2001encoding}
Daniel Gottesman, Alexei Kitaev, and John Preskill.
\newblock ``Encoding a qubit in an oscillator''.
\newblock Physical Review A {\bf 64}, 012310~(2001).
\newblock  url:~\url{https://doi.org/10.1103/PhysRevA.64.012310}.

\bibitem{michael2016new}
Marios~H Michael, Matti Silveri, RT~Brierley, Victor~V Albert, Juha Salmilehto,
  Liang Jiang, and Steven~M Girvin.
\newblock ``New class of quantum error-correcting codes for a bosonic mode''.
\newblock Physical Review X {\bf 6}, 031006~(2016).
\newblock  url:~\url{https://doi.org/10.1103/PhysRevX.6.031006}.

\bibitem{ofek2016extending}
Nissim Ofek, Andrei Petrenko, Reinier Heeres, Philip Reinhold, Zaki Leghtas,
  Brian Vlastakis, Yehan Liu, Luigi Frunzio, SM~Girvin, Liang Jiang, et~al.
\newblock ``Extending the lifetime of a quantum bit with error correction in
  superconducting circuits''.
\newblock Nature {\bf 536}, 441--445~(2016).
\newblock  url:~\url{https://doi.org/10.1038/nature18949}.

\bibitem{hu2019quantum}
Ling Hu, Yuwei Ma, Weizhou Cai, Xianghao Mu, Yuan Xu, Weiting Wang, Yukai Wu,
  Haiyan Wang, YP~Song, C-L Zou, et~al.
\newblock ``Quantum error correction and universal gate set operation on a
  binomial bosonic logical qubit''.
\newblock Nature Physics {\bf 15}, 503--508~(2019).
\newblock  url:~\url{https://doi.org/10.1038/s41567-018-0414-3}.

\bibitem{sivak2023real}
VV~Sivak, Alec Eickbusch, Baptiste Royer, Shraddha Singh, Ioannis Tsioutsios,
  Suhas Ganjam, Alessandro Miano, BL~Brock, AZ~Ding, Luigi Frunzio, et~al.
\newblock ``Real-time quantum error correction beyond break-even''.
\newblock Nature {\bf 616}, 50--55~(2023).
\newblock  url:~\url{https://doi.org/10.1038/s41586-023-05782-6}.

\bibitem{noh2022low}
Kyungjoo Noh, Christopher Chamberland, and Fernando~GSL Brand{\~a}o.
\newblock ``Low-overhead fault-tolerant quantum error correction with the
  surface-gkp code''.
\newblock PRX Quantum {\bf 3}, 010315~(2022).
\newblock  url:~\url{https://doi.org/10.1103/PRXQuantum.3.010315}.

\bibitem{chamberland2022building}
Christopher Chamberland, Kyungjoo Noh, Patricio Arrangoiz-Arriola, Earl~T
  Campbell, Connor~T Hann, Joseph Iverson, Harald Putterman, Thomas~C
  Bohdanowicz, Steven~T Flammia, Andrew Keller, et~al.
\newblock ``Building a fault-tolerant quantum computer using concatenated cat
  codes''.
\newblock PRX Quantum {\bf 3}, 010329~(2022).
\newblock  url:~\url{https://doi.org/10.1103/PRXQuantum.3.010329}.

\bibitem{li2017cat}
Linshu Li, Chang-Ling Zou, Victor~V Albert, Sreraman Muralidharan, SM~Girvin,
  and Liang Jiang.
\newblock ``Cat codes with optimal decoherence suppression for a lossy bosonic
  channel''.
\newblock Physical review letters {\bf 119}, 030502~(2017).
\newblock  url:~\url{https://doi.org/10.1103/PhysRevLett.119.030502}.

\bibitem{albert2018performance}
Victor~V Albert, Kyungjoo Noh, Kasper Duivenvoorden, Dylan~J Young,
  RT~Brierley, Philip Reinhold, Christophe Vuillot, Linshu Li, Chao Shen,
  SM~Girvin, et~al.
\newblock ``Performance and structure of single-mode bosonic codes''.
\newblock Physical Review A {\bf 97}, 032346~(2018).
\newblock  url:~\url{https://doi.org/10.1103/PhysRevA.97.032346}.

\bibitem{mirrahimi2014dynamically}
Mazyar Mirrahimi, Zaki Leghtas, Victor~V Albert, Steven Touzard, Robert~J
  Schoelkopf, Liang Jiang, and Michel~H Devoret.
\newblock ``Dynamically protected cat-qubits: a new paradigm for universal
  quantum computation''.
\newblock New Journal of Physics {\bf 16}, 045014~(2014).
\newblock  url:~\url{https://doi.org/10.1088/1367-2630/16/4/045014}.

\bibitem{leghtas2015confining}
Zaki Leghtas, Steven Touzard, Ioan~M Pop, Angela Kou, Brian Vlastakis, Andrei
  Petrenko, Katrina~M Sliwa, Anirudh Narla, Shyam Shankar, Michael~J Hatridge,
  et~al.
\newblock ``Confining the state of light to a quantum manifold by engineered
  two-photon loss''.
\newblock Science {\bf 347}, 853--857~(2015).
\newblock  url:~\url{https://doi.org/10.1126/science.aaa2085}.

\bibitem{lescanne2020exponential}
Rapha{\"e}l Lescanne, Marius Villiers, Th{\'e}au Peronnin, Alain Sarlette,
  Matthieu Delbecq, Benjamin Huard, Takis Kontos, Mazyar Mirrahimi, and Zaki
  Leghtas.
\newblock ``Exponential suppression of bit-flips in a qubit encoded in an
  oscillator''.
\newblock Nature Physics {\bf 16}, 509--513~(2020).
\newblock  url:~\url{https://doi.org/10.1038/s41567-020-0824-x}.

\bibitem{puri2017engineering}
Shruti Puri, Samuel Boutin, and Alexandre Blais.
\newblock ``Engineering the quantum states of light in a kerr-nonlinear
  resonator by two-photon driving''.
\newblock npj Quantum Information {\bf 3}, 18~(2017).
\newblock  url:~\url{https://doi.org/10.1038/s41534-017-0019-1}.

\bibitem{grimm2020stabilization}
Alexander Grimm, Nicholas~E Frattini, Shruti Puri, Shantanu~O Mundhada, Steven
  Touzard, Mazyar Mirrahimi, Steven~M Girvin, Shyam Shankar, and Michel~H
  Devoret.
\newblock ``Stabilization and operation of a kerr-cat qubit''.
\newblock Nature {\bf 584}, 205--209~(2020).
\newblock  url:~\url{https://doi.org/10.1038/s41586-020-2587-z}.

\bibitem{gautier2022combined}
Ronan Gautier, Alain Sarlette, and Mazyar Mirrahimi.
\newblock ``Combined dissipative and hamiltonian confinement of cat qubits''.
\newblock PRX Quantum {\bf 3}, 020339~(2022).
\newblock  url:~\url{https://doi.org/10.1103/PRXQuantum.3.020339}.

\bibitem{rosenblum2018fault}
Serge Rosenblum, Philip Reinhold, Mazyar Mirrahimi, Liang Jiang, Luigi Frunzio,
  and Robert~J Schoelkopf.
\newblock ``Fault-tolerant detection of a quantum error''.
\newblock Science {\bf 361}, 266--270~(2018).
\newblock  url:~\url{https://doi.org/10.1126/science.aat3996}.

\bibitem{puri2020bias}
Shruti Puri, Lucas St-Jean, Jonathan~A Gross, Alexander Grimm, Nicholas~E
  Frattini, Pavithran~S Iyer, Anirudh Krishna, Steven Touzard, Liang Jiang,
  Alexandre Blais, et~al.
\newblock ``Bias-preserving gates with stabilized cat qubits''.
\newblock Science advances {\bf 6}, eaay5901~(2020).
\newblock  url:~\url{https://doi.org/10.1126/sciadv.aay5901}.

\bibitem{xu2022engineering}
Qian Xu, Joseph~K Iverson, Fernando~GSL Brand{\~a}o, and Liang Jiang.
\newblock ``Engineering fast bias-preserving gates on stabilized cat qubits''.
\newblock Physical Review Research {\bf 4}, 013082~(2022).
\newblock  url:~\url{https://doi.org/10.1103/PhysRevResearch.4.013082}.

\bibitem{grimsmo2020quantum}
Arne~L Grimsmo, Joshua Combes, and Ben~Q Baragiola.
\newblock ``Quantum computing with rotation-symmetric bosonic codes''.
\newblock Physical Review X {\bf 10}, 011058~(2020).
\newblock  url:~\url{https://doi.org/10.1103/PhysRevX.10.011058}.

\bibitem{hillmann2022performance}
Timo Hillmann, Fernando Quijandr{\'\i}a, Arne~L Grimsmo, and Giulia Ferrini.
\newblock ``Performance of teleportation-based error-correction circuits for
  bosonic codes with noisy measurements''.
\newblock PRX Quantum {\bf 3}, 020334~(2022).
\newblock  url:~\url{https://doi.org/10.1103/PRXQuantum.3.020334}.

\bibitem{leghtas2013hardware}
Zaki Leghtas, Gerhard Kirchmair, Brian Vlastakis, Robert~J Schoelkopf, Michel~H
  Devoret, and Mazyar Mirrahimi.
\newblock ``Hardware-efficient autonomous quantum memory protection''.
\newblock Physical Review Letters {\bf 111}, 120501~(2013).
\newblock  url:~\url{https://doi.org/10.1103/PhysRevLett.111.120501}.

\bibitem{holevo2011probabilistic}
Alexander Holevo.
\newblock ``Probabilistic and statistical aspects of quantum theory''.
\newblock Springer Science. ~(2011).
\newblock  url:~\url{https://doi.org/10.1007/978-88-7642-378-9}.

\bibitem{wiseman1998adaptive}
Howard~M Wiseman and Rowan~B Killip.
\newblock ``Adaptive single-shot phase measurements: The full quantum theory''.
\newblock Physical Review A {\bf 57}, 2169~(1998).
\newblock  url:~\url{https://doi.org/10.1103/PhysRevA.57.2169}.

\bibitem{blais2021circuit}
Alexandre Blais, Arne~L Grimsmo, Steven~M Girvin, and Andreas Wallraff.
\newblock ``Circuit quantum electrodynamics''.
\newblock Reviews of Modern Physics {\bf 93}, 025005~(2021).
\newblock  url:~\url{https://doi.org/10.1103/RevModPhys.93.025005}.

\bibitem{aliferis2007level}
Panagiotis~(Panos) Aliferis.
\newblock ``Level reduction and the quantum threshold theorem''.
\newblock PhD thesis.
\newblock California Institute of Technology.
\newblock ~(2007).
\newblock  url:~\url{https://doi.org/10.48550/arXiv.quant-ph/0703230}.

\bibitem{Xu2024FToperation}
Qian Xu, Pei Zeng, Daohong Xu, and Liang Jiang.
\newblock ``Fault-tolerant operation of bosonic qubits with discrete-variable
  ancillae''.
\newblock Phys. Rev. X {\bf 14}, 031016~(2024).
\newblock  url:~\url{https://doi.org/10.1103/PhysRevX.14.031016}.

\bibitem{AGP2006}
Panos Aliferis, Daniel Gottesman, and John Preskill.
\newblock ``Quantum accuracy threshold for concatenated distance-3 codes''.
\newblock Quantum Info. Comput. {\bf 6}, 97~(2006).
\newblock  url:~\url{https://doi.org/10.48550/arXiv.quant-ph/0504218}.

\bibitem{joshi2021quantum}
Atharv Joshi, Kyungjoo Noh, and Yvonne~Y Gao.
\newblock ``Quantum information processing with bosonic qubits in circuit
  qed''.
\newblock Quantum Science and Technology {\bf 6}, 033001~(2021).
\newblock  url:~\url{https://doi.org/10.1088/2058-9565/abe989}.

\bibitem{schlegel2022quantum}
David~S Schlegel, Fabrizio Minganti, and Vincenzo Savona.
\newblock ``Quantum error correction using squeezed schr{\"o}dinger cat
  states''.
\newblock Physical Review A {\bf 106}, 022431~(2022).
\newblock  url:~\url{https://doi.org/10.1103/PhysRevA.106.022431}.

\bibitem{teh2020overcoming}
RY~Teh, PD~Drummond, and MD~Reid.
\newblock ``Overcoming decoherence of schr{\"o}dinger cat states formed in a
  cavity using squeezed-state inputs''.
\newblock Physical Review Research {\bf 2}, 043387~(2020).
\newblock  url:~\url{https://doi.org/10.1103/PhysRevResearch.2.043387}.

\bibitem{hillmann2023quantum}
Timo Hillmann and Fernando Quijandr{\'\i}a.
\newblock ``Quantum error correction with dissipatively stabilized squeezed-cat
  qubits''.
\newblock Physical Review A {\bf 107}, 032423~(2023).
\newblock  url:~\url{https://doi.org/10.1103/PhysRevA.107.032423}.

\bibitem{xu2023autonomous}
Qian Xu, Guo Zheng, Yu-Xin Wang, Peter Zoller, Aashish~A Clerk, and Liang
  Jiang.
\newblock ``Autonomous quantum error correction and fault-tolerant quantum
  computation with squeezed cat qubits''.
\newblock npj Quantum Information {\bf 9}, 78~(2023).
\newblock  url:~\url{https://doi.org/10.1038/s41534-023-00746-0}.

\bibitem{le2018slowing}
Hanna Le~Jeannic, Adrien Cavaill{\`e}s, Kun Huang, Radim Filip, and Julien
  Laurat.
\newblock ``Slowing quantum decoherence by squeezing in phase space''.
\newblock Physical Review Letters {\bf 120}, 073603~(2018).
\newblock  url:~\url{https://doi.org/10.1103/PhysRevLett.120.073603}.

\bibitem{pan2023protecting}
Xiaozhou Pan, Jonathan Schwinger, Ni-Ni Huang, Pengtao Song, Weipin Chua,
  Fumiya Hanamura, Atharv Joshi, Fernando Valadares, Radim Filip, and Yvonne~Y
  Gao.
\newblock ``Protecting the quantum interference of cat states by phase-space
  compression''.
\newblock Physical Review X {\bf 13}, 021004~(2023).
\newblock  url:~\url{https://doi.org/10.1103/PhysRevX.13.021004}.

\bibitem{Matsuura2024CVFT}
Takaya Matsuura, Nicolas~C. Menicucci, and Hayata Yamasaki.
\newblock ``Continuous-variable fault-tolerant quantum computation under
  general noise''~(2024).
\newblock  url:~\url{https://doi.org/10.48550/arXiv.2410.12365}.

\bibitem{knill1997theory}
Emanuel Knill and Raymond Laflamme.
\newblock ``Theory of quantum error-correcting codes''.
\newblock Physical Review A {\bf 55}, 900~(1997).
\newblock  url:~\url{https://doi.org/10.1103/PhysRevA.55.900}.

\bibitem{bender2013advanced}
Carl~M Bender and Steven~A Orszag.
\newblock ``Advanced mathematical methods for scientists and engineers i:
  Asymptotic methods and perturbation theory''.
\newblock Springer Science \& Business Media. ~(2013).
\newblock  url:~\url{https://doi.org/10.1007/978-1-4757-3069-2}.

\bibitem{Petz1986sufficient}
Dénes Petz.
\newblock ``Sufficient subalgebras and the relative entropy of states of a von
  neumann algebra''.
\newblock Communications in Mathematical Physics {\bf 105}, 123--131~(1986).
\newblock  url:~\url{https://doi.org/10.1007/BF01212345}.

\bibitem{NgMandayam2010}
Hui~Khoon Ng and Prabha Mandayam.
\newblock ``Simple approach to approximate quantum error correction based on
  the transpose channel''.
\newblock Phys. Rev. A {\bf 81}, 062342~(2010).
\newblock  url:~\url{https://doi.org/10.1103/PhysRevA.81.062342}.

\bibitem{Horodeckis1999general}
Micha\l{} Horodecki, Pawe\l{} Horodecki, and Ryszard Horodecki.
\newblock ``General teleportation channel, singlet fraction, and
  quasidistillation''.
\newblock Physical Review A {\bf 60}, 1888--1898~(1999).
\newblock  url:~\url{https://doi.org/10.1103/PhysRevA.60.1888}.

\end{thebibliography}

\onecolumn
\appendix
\clearpage

\section{Correctability of cat codes}
\label{app:knill_laflamme}

Here, using the Knill-Laflamme error correction conditions \cite{knill1997theory}, we show that the order-$N$ cat code can simultaneously correct up to $N-1$ loss errors and phase errors smaller than $\frac{\pi}{N}$. Specifically, the set of correctable errors can be chosen to be [as in Eq.~\eqref{eq:QEC_E}],
\begin{align}
	&\cE \equiv \{E_{k,\theta}\equiv R(\theta)\widehat{a}^k\}\\
	\textrm{with } &k\textrm{ nonnegative integer}\in[k_0,k_0+N)\nonumber\\
	\textrm{and } &\theta \textrm{ real}\in(\theta_0,\theta_0+\tfrac{\pi}{N}].\nonumber
\end{align}   
The Knill-Laflamme conditions can be stated in terms of the logical cat-code states as
\begin{align}
	\bra{0_{N,\alpha}}E_{k,\theta}^\dagger E_{k',\theta'}\ket{0_{N,\alpha}}&=\bra{1_{N,\alpha}}E_{k,\theta}^\dagger E_{k',\theta'}\ket{1_{N,\alpha}}\nonumber\\
	\textrm{and}~~ \bra{0_{N,\alpha}}E_{k,\theta}^\dagger E_{k',\theta'}\ket{1_{N,\alpha}}&=0,
\end{align}
for all $E_{k,\theta},E_{k',\theta'}\in\cE$. $\cE$ is a correctable set of errors for the order-$N$ cat code if these conditions are satisfied.

To check these conditions, we begin with the expressions for $\ket{0_{N,\alpha}}$ and $\ket{1_{N,\alpha}}$ in the Fock basis [Eq.~\eqref{eq:cat_code_fock}]:
\begin{equation}
	\ket{\mu_{N, \alpha}} = \frac{2N\upe^{-\alpha^2/2}}{\sqrt{\cN_\mu}}\!\!\sum_{m\in\mathbb{Z}_\mu} \!\!\frac{\alpha^{mN}}{\sqrt{(mN)!}} \ket{mN},~ \mu=0,1.
\end{equation}
The action of $E_{k,\theta}\in\cE$ on these logical states is
\begin{align}
	E_{k,\theta}\!\ket{\mu_{N,\alpha}}\!=\!\frac{2N\upe^{-\alpha^2/2}}{\sqrt{\cN_\mu}}\upe^{-\upi k\theta}\!\!\!\sum_{m\in\mathbb{Z}_\mu} \!\!\!\frac{(\upe^{\upi\theta}\alpha)^{mN}}{\sqrt{(mN\!\!-\!\!k)!}} \ket{mN\!-\!k}\!,
\end{align}
where the sum over $m$ is only over those $m$ with $mN-k>0$. Recalling the definition of $\mathbb{Z}_\mu$, we see that $E_{k,\theta}\ket{0_{N,\alpha}}$ is supported only on the $\ket{(\textrm{even})N-k}$ Fock states, while $E_{k,\theta}\ket{0_{N,\alpha}}$ is supported only on the $\ket{(\textrm{odd})N-k}$ Fock states. Due to these disjoint supports, we have that
\begin{align}
	\bra{0_{N,\alpha}}E_{k,\theta}^\dagger E_{k',\theta'}\ket{1_{N,\alpha}}=0,
\end{align}
for any integers $E_{k,\theta},E_{k'\theta'}\in\cE$. In addition, when $k\neq k'$, we have,
\begin{align}
	\bra{\mu_{N,\alpha}}E_{k,\theta}^\dagger E_{k',\theta'}\ket{\mu_{N,\alpha}}=0,~~\mu=0,1.
\end{align}
All that remains is to check the case for $k=k'$. Straightforward algebra gives, for $\mu=0,1$,
\begin{align}\label{eq:Emu}
	&\quad \bra{\mu_{N,\alpha}}E_{k,\theta}^\dagger E_{k,\theta'}\ket{\mu_{N,\alpha}}=\frac{\alpha^{2k}}{\cN_\mu}2N{\left[\upe^{-\alpha^2{\left(1-\upe^{-\upi(\theta-\theta')}\right)}}+F_\mu(k;\theta,\theta')\right]},\nonumber
\end{align}
with
\begin{align}
	F_\mu(k;\theta,\theta')&\equiv \sum_{m=1}^{2N-1}(-1)^{\mu m}\upe^{-\upi\frac{\pi}{N}km} \times\exp{-\alpha^2{\left[1-\upe^{-\upi\frac{\pi}{N}m}\upe^{-\upi(\theta-\theta')}\right]}}.\nonumber
\end{align}
Using the fact that $|\sum x|\leq \sum|x|$, one can show that
\begin{equation}
	|F_\mu(k;\theta,\theta')|\leq (2N-1)\upe^{-\alpha^2[1-c(\theta,\theta')]},
\end{equation}
with $c(\theta,\theta')\equiv \max_{m\in\{1,2\dots,2N-1\}}\{\cos[\tfrac{\pi m}{N}+(\theta-\theta')]\}$, vanishingly small when $\alpha\rightarrow \infty$. Note that the normalization constants $\cN_\mu$ are given by
\begin{equation}
	\cN_\mu=2N[1+F_\mu(0;0,0)].
\end{equation}
Observe that, for $N\sim O(1)$ and large $\alpha$, $F_\mu(0;0,0)$ is a small correction to the ``$1$'' term in $\cN_\mu$, so that $\cN_0\simeq \cN_1$. Using Eq.~\eqref{eq:Emu}, we have
\begin{align}\label{eq:kl_violation}
	&\quad \bra{0_{N,\alpha}}E_{k,\theta}^\dagger E_{k,\theta'}\ket{0_{N,\alpha}}-\bra{1_{N,\alpha}}E_{k,\theta}^\dagger E_{k,\theta'}\ket{1_{N,\alpha}}\nonumber\\
	&=2N\alpha^{2k}{\left\{\upe^{-\alpha^2(1-\upe^{-\upi(\theta-\theta')})}{\left[\frac{1}{\cN_0}-\frac{1}{\cN_1}\right]}
	+\frac{F_0(k;\theta,\theta')}{\cN_0}-\frac{F_1(k;\theta,\theta')}{\cN_1}\right\}}.\nonumber
\end{align}

Now, we note that
\begin{align}
	\frac{1}{\cN_0}-\frac{1}{\cN_1} &= \frac{F_1(0;0,0)-F_0(0;0,0)}{2N(1+F_0(0;0,0))(1+F_1(0;0,0))}\\
	&\leq \frac{1}{2N}[|F_1(0;0,0)|+|F_0(0;0,0)|] \leq \frac{2N-1}{N}\upe^{-\alpha^2[1-c(0,0)]},\nonumber
\end{align} 
vanishingly small for large $\alpha$. Similarly, we have
\begin{align}
	\tfrac{1}{\cN_0}F_0(k;\theta,\theta')-\tfrac{1}{\cN_1}F_1(k;\theta,\theta') &\leq \tfrac{1}{2N}[|F_0(k;\theta,\theta')| (1+|F_1(0;0,0)|) +|F_1(k;\theta,\theta')|(1+|F_0(0;0,0)|)] \\
	&\leq \tfrac{2N-1}{N}\upe^{-\alpha^2[1-c(\theta,\theta')]} (1+\upe^{-\alpha^2[1-c(0,0)]}).\nonumber
\end{align}
Since $c(0,0) < 1$ and $c(\theta, \theta')<1$ when $\theta, \theta' \in [\theta_0+\epsilon,\theta_0+\pi/N]$ for any finite $\epsilon$, the quantity in Eq.~\ref{eq:kl_violation} is vanishingly small when $\alpha\rightarrow\infty$. Overall, the cat code and the set of errors $\cE$ only approximately satisfy the Knill-Laflamme conditions. However, the amount of violation decreases exponentially with the squared amplitude $\alpha^2$.

\section{Errors in the discrete phase measurement}
\label{app:phase_meas}

Here, we analyze the measurement error inherent in the discrete phase measurement introduced in Sec.~\ref{sec:meas}. As defined earlier, the discrete phase measurement has measurement operators [Eq.~\eqref{eq:DPmeas}]
\begin{equation}
	\Pi^{\textrm{DP}}_K(k;\phi_0)\equiv \int_{{\left(k-\tfrac{1}{2}\right)}\tfrac{2\pi}{K}+\phi_0}^{{\left(k+\tfrac{1}{2}\right)}\tfrac{2\pi}{K}+\phi_0} \upd\phi~\Pi(\phi), \quad k=0,1,\ldots, K,
\end{equation}
for $k\in\{0,1,2,\ldots, K\}$, used to estimate the value of $k$ for the rotated cat-code state $R{\left(k\frac{2\pi}{K}+\phi_0\right)}\ket{+_{N,\alpha}}$, for a known offset $\phi_0$. Here, $\Pi(\phi)\equiv \ket{\phi}\bra{\phi}$, with $\ket{\phi}\equiv \frac{1}{\sqrt{2\pi}}\sum_{n=0}^\infty\upe^{\upi n\phi}\ket{n}$, the phase states defined earlier.

An error in the measurement occurs when, for the state $R{\left(k\frac{2\pi}{K}+\phi_0\right)}\ket{+_{N,\alpha}}$, a $k'\neq k$ outcome is obtained from the discrete phase measurement. The error probability is given by, for $k'\neq k$ (with $\phi_0=0$ for brevity, as it does not affect the answer),
\begin{align} \label{eq:phase_error}
	p_\textrm{err}&\equiv \bra{+_{N,\alpha}}R{\left(k\tfrac{2\pi}{K}\right)}^\dagger\Pi^{\textrm{DP}}_K(k';0)R{\left(k\tfrac{2\pi}{K}\right)}\ket{+_{N,\alpha}}\\
	&=\int_{{\left(k'-\tfrac{1}{2}\right)}\tfrac{2\pi}{K}}^{{\left(k'+\tfrac{1}{2}\right)}\tfrac{2\pi}{K}} \upd\phi~{\left|\bra{\phi}R{\left(k\tfrac{2\pi}{K}\right)}\ket{+_{N,\alpha}}\right|}^2.\nonumber
\end{align}
We assume that $\alpha$ is large enough so that we can neglect the difference between the normalization constants and set $\cN_0\simeq\cN_1\simeq2N$. Then, straightforward algebra gives
\begin{equation}
	{\left|\!\bra{\phi}R{\left(k\tfrac{2\pi}{K}\right)}\ket{+_{N,\alpha}}\!\right|}^2\!\!=\!\tfrac{N}{2\pi}\upe^{-\alpha^2}\!\!\!\sum_{\ell,\ell'=0}^\infty\!\!\!\frac{\alpha^{(\ell+\ell')N}\upe^{\upi(\ell-\ell')N{\left(\!\frac{2\pi k}{K}-\phi\!\right)}}}{\sqrt{(\ell N)!(\ell'N)!}}.
\end{equation}
We need to integrate this expression over $\phi$. For the $\ell=\ell'$ terms in the sum, we have
\begin{align}\label{eq:diag_terms}
	&\quad \int_{{\left(k'-\tfrac{1}{2}\right)}\tfrac{2\pi}{K}}^{{\left(k'+\tfrac{1}{2}\right)}\tfrac{2\pi}{K}}\upd\phi~\tfrac{N}{2\pi}\upe^{-\alpha^2}\sum_{\ell=0}^\infty\frac{\alpha^{2\ell N}}{(\ell N)!}\\
	&=\tfrac{N}{K}\upe^{-\alpha^2}\sum_{\ell=0}^\infty\frac{\alpha^{2\ell N}}{(\ell N)!}=\tfrac{1}{K}\upe^{-\alpha^2}\sum_{n=0}^\infty\sum_{s=0}^{N-1}\frac{\alpha^{2n}}{n!}\upe^{\upi\frac{2\pi}{N}ns}\nonumber\\
	&=\tfrac{1}{K}\upe^{-\alpha^2}\!\!\sum_{s=0}^{N-1}\upe^{\alpha^2\upe^{\upi\frac{2\pi s}{N}}}\!\!\!= \tfrac{1}{K}{\left[1 + \upe^{-\alpha^2}\!\!\sum_{s=1}^{N-1}\frac{\upe^{\alpha^2\upe^{\upi\frac{2\pi s}{N}}}\!\! \!+ \upe^{\alpha^2\upe^{-\upi\frac{2\pi s}{N}}}}{2}\right]}\nonumber\\
	&= \tfrac{1}{K}\left(1 + \sum_{s=1}^{N-1} \upe^{-\alpha^2(1 - \cos{\frac{2\pi s}{N}})} \cos{(\alpha^2\sin{\frac{2\pi s}{N}})}\right)\nonumber\\
	&=\tfrac{1}{K}{\left[1+O(\upe^{-c\alpha^2})\right]},\nonumber
\end{align}
for $c\equiv 1-\cos{\left(\frac{2\pi}{N}\right)}$. The $\ell\neq\ell'$ terms evaluate to
\begin{align} \label{eq:off_diag}
	&\quad\tfrac{1}{\pi}\upe^{-\alpha^2}\sum_{\ell\neq\ell'}\frac{\alpha^{(\ell+\ell')N}\upe^{\upi(\ell-\ell')N(k-k')\frac{2\pi}{K}}}{\sqrt{(\ell N)!(\ell'N)!}}\frac{\sin{\left[(\ell-\ell')\frac{\pi N}{K}\right]}}{\ell-\ell'}.
\end{align}

We specialize to the case of the $\overline X$ measurement by setting $K=2N$ [see Eq.~\eqref{eq:x_meas}]. An error occurs if the outcome $k'$ and the true value $k$ have different parities. We set $k-k'=1$, which is all we need for the $\overline X$ measurement.

We want to show that the quantity in Eq.~\eqref{eq:off_diag} behaves as $-\frac{1}{2N}{\left[1 + O(\upe^{-c'\alpha^2})\right]}$ for some constant $c'$ in the limit $\alpha \rightarrow \infty$. Combined with Eq.~\eqref{eq:diag_terms}, this would mean that the error probability for the $\overline X$ measurement is exponentially suppressed in $\alpha^2$.  

We first simplify the sum in Eq.~\eqref{eq:off_diag} by observing that a term is non-zero only if $\ell - \ell'$ is an odd number. Then,
\begin{align}
	\text{(32)} &= \tfrac{1}{\pi}\upe^{-\alpha^2}\sum_{\ell > \ell'}\frac{\alpha^{(\ell+\ell')N}\upe^{\upi(\ell-\ell')\pi}}{\sqrt{(\ell N)!(\ell'N)!}}\frac{\sin{\left[(\ell-\ell')\frac{\pi}{2}\right]}}{\ell-\ell'} 
	+ \tfrac{1}{\pi}\upe^{-\alpha^2}\sum_{\ell<\ell'}\frac{\alpha^{(\ell+\ell')N}\upe^{\upi(\ell-\ell')\pi}}{\sqrt{(\ell N)!(\ell'N)!}}\frac{\sin{\left[(\ell-\ell')\frac{\pi}{2}\right]}}{\ell-\ell'} \\
	&= \tfrac{2}{\pi}\upe^{-\alpha^2}\sum_{\ell > \ell'}\frac{\alpha^{(\ell+\ell')N}\cos{((\ell-\ell')\pi)}}{\sqrt{(\ell N)!(\ell'N)!}}\frac{\sin{\left[(\ell-\ell')\frac{\pi}{2}\right]}}{\ell-\ell'} \nonumber\\
	&= -\tfrac{2}{\pi}\upe^{-\alpha^2}\sum_{\ell > \ell'}\frac{\alpha^{(\ell+\ell')N}}{\sqrt{(\ell N)!(\ell'N)!}}\frac{\sin{\left[(\ell-\ell')\frac{\pi}{2}\right]}}{\ell-\ell'},\nonumber
\end{align}
where, in the last equality, we have used the fact that $\ell-\ell'$ is odd. 
Letting $\ell + \ell' = 2r+1$ and $\ell - \ell' = 2p+1$, we have,
\begin{align}
	\label{eq:off_diag_2}
	-\frac{2}{\pi}\upe^{-\alpha^2}\sum_{p=0}^{\infty}\sum_{r=p}^{\infty} \frac{\alpha^{(2r+1)N}}{\sqrt{((r+p+1)N)!((r-p)N)!}}\frac{(-1)^p}{2p+1} = -\frac{2}{\pi N}\upe^{-\alpha^2}\sum_{s=0}^{N-1}\sum_{p=0}^{\infty}\frac{(-1)^p}{2p+1} \alpha^{(2p+1)N} G^{(s)}_p(\alpha),
\end{align}
where 
\begin{align}
	G^{(s)}_p(\alpha) = \sum_{n=0}^{\infty} \frac{\alpha^{2n}\cos{\frac{2\pi sn}{N}}}{\sqrt{n!(n+m_p)!}}.
\end{align}
Here, $m_p \equiv (2p+1)N$. In the second step, we change variables $r \rightarrow r-p$ and used the identity $\frac{1}{N}\sum_{s=0}^{N-1} \upe^{\pm\upi \tfrac{2\pi n}{N}s} = \sum_{r=0}^{\infty} \delta_{n, rN}$.

Now, we want to approximate the asymptotic behavior of $G^{(s)}_p(\alpha)$ when $\alpha \rightarrow \infty$, following the method described in Ref.~\cite{bender2013advanced}. We notice that the series $G^{(s)}_p(\alpha)$ is a discrete cosine function modulated by $g_n \equiv \frac{\alpha^{2n}}{\sqrt{n!(n+m_p)!}}$. The ratio of two consecutive terms is $\frac{g_{n}}{g_{n-1}} = \frac{\alpha^2}{\sqrt{n(n+m_p)}}$, for $n=1,2,\ldots$. The summands thus peak around the term with index $n_0$ such that $g_{n_0}/g_{n_0-1}=1$, or $n_0 = \frac{1}{2}{\left(-m_p + \sqrt{m_p^2 + 4\alpha^4}\right)}$. $G^{(s)}_p(\alpha)$ can thus be approximated by summing around $n_0$. Specifically, for a given $\varepsilon > 0$, 
\begin{align}\label{eq:gp}
	G^{(s)}_p(\alpha) \approx \sum_{n=[n_0(1-\varepsilon)]}^{n_0(1+\varepsilon)} \frac{\alpha^{2n} \cos{\frac{2\pi sn}{N}} }{\sqrt{n!((n+m_p)!)}}.
\end{align}
In this region, writing $n = n_0 + t$ ($t \ll n_0$) and using Stirling's formula, we can approximate $n!$ and $(n+m_p)!$:
\begin{align*}
	n! & \approx \sqrt{2\pi n_0} \upe^{n_0\ln(n_0) - n_0} \upe^{t\ln(n_0)+\tfrac{t^2}{2n_0}}, \\
	(n+m_p)! &\approx \sqrt{2\pi (n_0+m_p)} \upe^{(n_0+m_p)\ln(n_0+m_p) - (n_0+m_p)}  \times\upe^{t\ln(n_0+m_p)+\tfrac{t^2}{2(n_0+m_p)}}.\nonumber
\end{align*}
Combining this with the fact that $n_0(n_0+m_p)=\alpha^4$, we have
\begin{align*}
	\frac{1}{\sqrt{n!(n+m_p)!}} &\approx \frac{1}{\sqrt{2\pi}} \frac{\upe^{n_0 + m_p/2}}{\sqrt{ (n_0+m_p)^{m_p}}}
	\times \alpha^{-(2n_0+2t+1)} \upe^{-\tfrac{2n_0+m_p}{4\alpha^4}t^2}.\nonumber
\end{align*}
Substituting this in Eq.~\eqref{eq:gp} and extending the sum to infinity, we arrive at 
\begin{align*}
	G^{(s)}_p(\alpha) \approx \sum_{t=-\infty}^{\infty} &\frac{1}{\sqrt{2\pi\alpha^2}} \frac{\upe^{n_0 + m_p/2}}{\sqrt{ (n_0+m_p)^{m_p}}} 
	\times \upe^{-\tfrac{2n_0+m_p}{4\alpha^4}t^2} \cos{\frac{2\pi s (n_0 + t)}{N}}.
\end{align*}
As $\alpha\rightarrow\infty$, we can approximate this sum as the Riemann sum for the following integral,
\begin{align*}
	G^{(s)}_p(\alpha) &\approx \frac{1}{\sqrt{2\pi}} \frac{\upe^{n_0 + m_p/2}}{\sqrt{ (n_0+m_p)^{m_p}}} \times\int_{-\infty}^{\infty} \mathrm{d}x\,\upe^{-\tfrac{2n_0+m_p}{4\alpha^2}x^2} \cos{\frac{2\pi s (n_0 + \alpha x)}{N}} \\
	&= \frac{1}{\sqrt{2\pi}} \frac{\upe^{n_0 + m_p/2}}{\sqrt{ (n_0+m_p)^{m_p}}}
	\times \frac{1}{2}\int_{-\infty}^{\infty} \mathrm{d}x\,\upe^{-\tfrac{2n_0+m_p}{4\alpha^2}x^2} \left(\upe^{\upi\frac{2\pi s}{N}(n_0 + \alpha x)} + \upe^{-\upi\frac{2\pi s}{N} (n_0 + \alpha x)}\right)
	\\
	&= \frac{1}{\sqrt{2\pi}} \frac{\upe^{n_0 + m_p/2}}{\sqrt{ (n_0+m_p)^{m_p}}} \sqrt{\frac{4\pi \alpha^2}{2n_0 + m_p}} \upe^{-\frac{4\pi^2s^2\alpha^4}{(2n_0+m_p)N^2}} \cos{\frac{2\pi sn_0}{N}}.
\end{align*}
In the last step, we used the Gaussian integral $\int_{-\infty}^{\infty}\mathrm{d}x \upe^{-\frac{1}{2}ax^2+\upi bx} = \sqrt{\frac{2\pi}{a}}\upe^{-\frac{b^2}{2a}}$. Let $x_p \equiv \frac{m_p}{2\alpha^2}$ and substitute the expression for $n_0$. We can then write $G^{(s)}_p(\alpha)$ as
\begin{align}
	\label{eq:gps_approx}
	G^{(s)}_p(\alpha) &\approx \alpha^{-m_p}(1+x_p)^{-1/4} \cos(\frac{2\pi s\alpha^2}{N}(\sqrt{x_p^2+1}-x_p)) \nonumber\\ 
	&\times\upe^{\alpha^2 \left(\sqrt{1+ x_p^2} - x_p\ln(x_p + \sqrt{1+x_p^2}) - \frac{2\pi^2 s^2}{N^2\sqrt{1+x_p^2}}\right)}.
\end{align}
The error introduced in the series of approximations above is a decreasing function of $\alpha$. Thus, with the factor of $\upe^{-\alpha^2}$ [see Eq.~\eqref{eq:off_diag_2}], it decays faster than $\upe^{-\alpha^2}$.

The next step is to notice that for $N>s>0$, the exponent of the exponential function in Eq.~\eqref{eq:gps_approx} is less than $\alpha^2$. Therefore, when combining with the factor of $\upe^{-\alpha^2}$, the sums corresponding to $N>s>0$ in Eq.~\ref{eq:off_diag_2} always decay as $O(\upe^{-c\alpha^2})$ for some constant $c$. Specifically, the function 
\[
f_s(x) = \sqrt{1+ x^2} - x\ln(x + \sqrt{1+x^2}) - \frac{2\pi^2 s^2}{N^2\sqrt{1+x^2}}
\]
peaks at $x_s>0$ satisfying $f_s'(x_s)=0$, or \[\ln(x_s + \sqrt{1+x_s^2}) = \frac{2\pi^2s^2}{N^2}\frac{x_s}{(1+x_s^2)^{3/2}}.\] The function \[f_s(x_s) = \sqrt{1+ x_s^2} - \left(\frac{1}{x_s}+2x_s\right) \ln(x_s + \sqrt{1+x_s^2})\] is a decreasing function for $x_s>0$. Therefore, $f_s(x) \leq f_s(x_s) < \lim_{x_s\rightarrow0}f_s(x_s) = 1$.

We focus now on the case $s=0$, for which 
\begin{align*}
	G^{(0)}_p(\alpha) \approx \alpha^{-m_p}(1+x_p)^{-1/4} 
	\upe^{\alpha^2 \left(\sqrt{1+ x_p^2} - x_p\ln(x_p + \sqrt{1+x_p^2})\right)}.
\end{align*}
The function $f_0(x) = \sqrt{1+ x^2} - x\ln(x + \sqrt{1+x^2})$ is a decreasing function for $x\geq 0$, since $f_0'(x) = -\ln(x + \sqrt{1+x^2}) \leq 0$. Thus, $G^{(0)}_p(\alpha)$ decreases with increasing $p$. Define $I(\alpha) \equiv \sum_{p=0}^{\infty} \frac{(-1)^p}{2p+1} \alpha^{(2p+1)N} G^{(0)}_p(\alpha)$. Using the same technique for approximating $G^{(s)}_p(\alpha)$, we can approximate $I_s(\alpha)$ by summing over terms around $p=0$ and expanding $f(x_p)$ around $x_0$. We have
\begin{align}
	I(\alpha) &\approx \sum_{p=0}^{\varepsilon} \frac{(-1)^p}{2p+1}
	(1+x_p)^{-1/4} \upe^{\alpha^2 \left(\sqrt{1+ x_p^2} - x_p\ln(x_p + \sqrt{1+x_p^2})\right)} \nonumber\\
	&\approx \sum_{p=0}^{\varepsilon} \frac{(-1)^p}{2p+1} \upe^{\alpha^2\left(1-\tfrac{(2p+1)^2N^2}{8\alpha^4}\right)} \nonumber\\
	&\approx \upe^{\alpha^2} \sum_{p=0}^{\infty} \frac{\cos(\pi p)}{2p+1} \upe^{-\tfrac{(2p+1)^2 N^2}{8\alpha^2}}.
\end{align}
Letting $t_p \equiv \frac{(2p+1)N}{2\alpha}$, we can evaluate $I(\alpha)$ as a Riemann sum,
\begin{align}
	&I(\alpha) \approx \upe^{\alpha^2} \sum_{p=0}^{\infty} \frac{N}{2\alpha}\frac{\cos(\tfrac{\pi\alpha t_p}{N}-\tfrac{\pi}{2})}{t_p} \upe^{-\tfrac{1}{2}t_p^2} \\
	&\approx \frac{1}{2}\upe^{\alpha^2}\!\!\! \int_{0}^{\infty} \!\!\!\mathrm{d}t \frac{\sin(\tfrac{\pi\alpha t}{N})}{t} \upe^{-t^2/2}= \frac{1}{2}\upe^{\alpha^2}\!\! \sqrt{\frac{\pi}{2}}\! \int_{0}^{\pi\alpha/N}\!\!\! \mathrm{d}x \,\upe^{-x^2/2}.\nonumber
\end{align}
In the limit of $\alpha\rightarrow \infty$, the last integral goes to $\sqrt{\tfrac{\pi}{2}}$, so that $I(\alpha) \approx \frac{\pi}{4}\upe^{\alpha^2}$ with an error suppressed exponentially in $\alpha^2$. Therefore, $I(\alpha)$ contributes a term $-\frac{1}{2N}$ to the sum in Eq.~\eqref{eq:off_diag_2}.

Altogether, the off-diagonal sum ($\ell \neq \ell'$) in Eq.~\eqref{eq:off_diag_2} behaves as $-\frac{1}{2N} + O(\upe^{-c\alpha^2})$ for some constant $c$. Together with the diagonal term [Eq.~\eqref{eq:diag_terms} with $K=2N$], the measurement error $p_\textrm{err}$ for the $\overline X$ measurement thus scales as $O(\upe^{-c\alpha^2})$ for large $\alpha$.

\section{Fault tolerance of the Knill and hybrid EC gadgets}
\label{app:ft}
Here, we show that the Knill and hybrid EC gadgets satisfy EC-FT property of Sec.~\ref{sec:ft_analysis}, and discuss the consequences on the code order for the ancillary mode. 

\subsection{Knill EC gadget}
\label{app:ft_knill}

\begin{figure}[h]
	\centering
	\includegraphics[trim=0mm 100mm 90mm 0mm, clip, width=0.45\columnwidth]{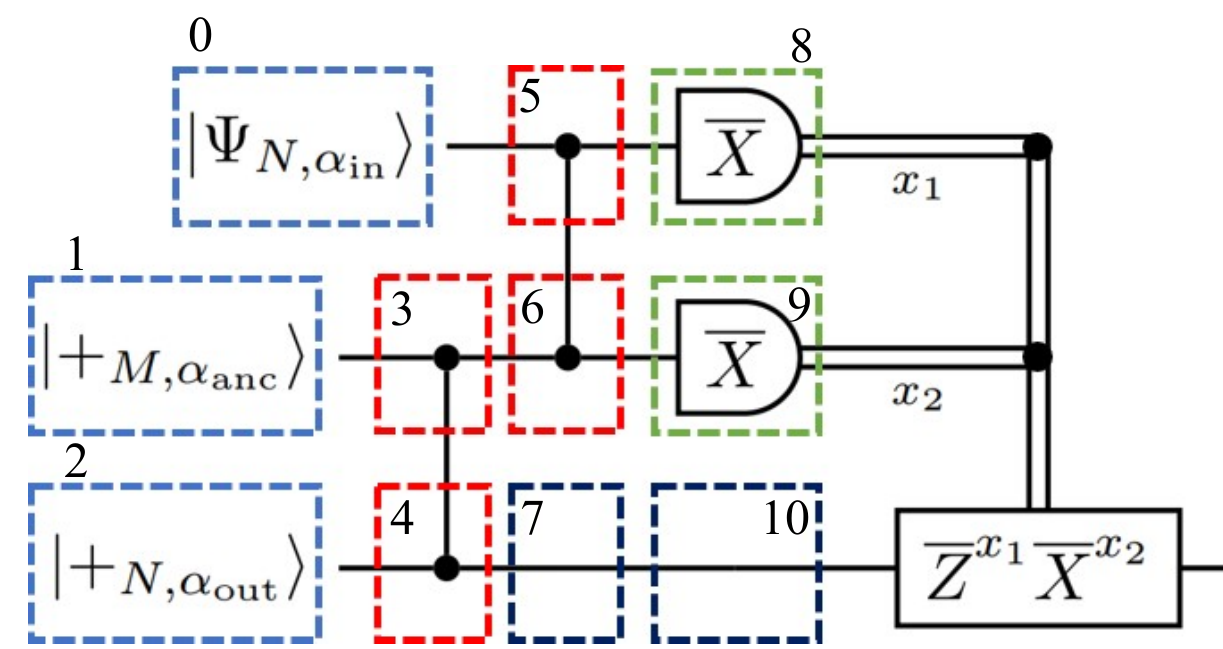}
	\caption{Fault locations in the Knill EC gadget.}
	\label{fig:knillFT}
\end{figure}

The Knill EC gadget comprises 10 locations labeled by $i\in\{1,2,\dots ,10\}$, as marked by colored boxes in Fig.~\ref{fig:knillFT}, which includes two waiting locations in the output mode. Note that, in our noise model as described in Sec.~\ref{sec:noise}, a two-mode gate is marked as two separate locations as we assume the noise is applied on the two participating modes independently (see the Conclusions section in the main text for further discussion of this point). Additionally, there is one input location labeled as $0$. A location corresponds to a place in the circuit where a fault can occur, i.e., something goes wrong, leading to errors in the modes contained in that location. Note that the final classically controlled $\overline Z^a\overline X^b$ gate is not marked as a location: The recovery is implemented as a virtual (classical) Pauli-frame rotation that requires no actual gate operation.

A fault in each location can lead to both loss and phase errors on the mode; we denote the number of losses in location $i$ as $k_i$, and the phase error is denoted as $\theta_i$. In inserting faults, we follow the same convention used in the simulations (see Sec.~\ref{sec:noise}): The noisy preparation of a state is modeled as the ideal preparation followed by an application of the fault; the noisy gate is modeled as the ideal gate followed by the fault; the noisy measurement is modeled as the ideal measurement \emph{preceded} by the fault.

From the error propagation properties of the $\overline{\text{CZ}}$ gate (see Fig.~\ref{fig:errorCROT}) and the commutation of loss and phase errors (up to a phase), we note that the errors at the end of the circuit for each mode, just before the final ideal measurements (but after any measurement faults), can be combined into the form $\widehat{a}^k\upe^{\upi\theta\widehat{n}}$ for some $k$ and $\theta$. We label those overall $k$ and $\theta$ values for the input (top), ancillary (middle), and output (bottom) modes with superscripts $(I)$, $(A)$, and $(O)$, respectively. For the moment, we will assume that $M=N$, i.e., the code for the ancillary mode is of the same order as that for the input and output modes; we explain what happens when $M\neq N$ at the end. Then, we have, 
\begin{align}\label{eq:KnillEC1}
	k^{(I)} &= k_0+k_5 + k_8\,,\\
	\theta^{(I)} &= \theta_0+\theta_5 + \theta_8 -\tfrac{\pi}{N^2}(k_1+k_3)\,,\nonumber\\
	k^{(A)} &= k_1 + k_3 + k_6 + k_9\,,\nonumber\\
	\theta^{(A)} &= \theta_1+ \theta_3  +\theta_6 + \theta_9 - \tfrac{\pi}{N^2}(k_2+k_0)\,,\nonumber\\
	k^{(O)} &= k_2 +k_4+ k_7 + k_{10},\nonumber\\
	\theta^{(O)} &= \theta_2 +\theta_4+ \theta_7 + \theta_{10} - \tfrac{\pi k_1}{N^2}.\nonumber
\end{align}
We recall that all angles $\theta_i$ are negative [$\in(-\pi/N,0]$] as assumed earlier (see the sentences before the statement of the EC-FT property). The phase propagated by $\crot(\varphi)$ from a $k$-loss error is $-k\varphi$ (see Fig.~\ref{fig:errorCROT}). The phase errors above thus all accumulate in the same direction.

We let $r\equiv\sum_{i=1}^{10} k_i$ and $\theta_f \equiv \sum_{\theta=1}^{10}\theta_i$ be the total number of losses and total phase errors, respectively, due to faults inside the gadget. Furthermore, we write $\theta_s=-\frac{\pi k_0}{N^2}$, and $\theta_r=-\frac{\pi}{N^2}(2k_1 +k_2+k_3)$ as the phase errors induced by losses in the input and the gadget, respectively. 
Assume that $0 \leq k_0+r < N$ and $|\theta_f + \theta_s + \theta_r |< \frac{\pi}{N}$. It is evident that for mode $n=I,A, O$, $k^{(n)}<N$ and $|\theta^{(n)}| < \frac{\pi}{N}$. In particular, $k^{(O)} \leq r$ and $|\theta^{(O)}| < |\theta_r + \theta_f|$, from the definition of $r$, $\theta_r$, and $\theta_s$. It means that the errors in the output mode are solely due to faults in the EC but not errors in the input mode. Moreover, since $|\theta^{(n)}| < \frac{\pi}{N}$, within the correction capacity of an order-$N$ code, the outcome of the $\overline{X}$ measurements will not be affected. As a consequence, there is no logical error in the output state and ideally decoding it should give the same state as ideally decoding the input state. Altogether then, the Knill EC gadget satisfies the EC-FT property.

In the previous discussion, we assumed that $N=M$ for simplicity. Let us now relax that constraint. If $N>M$, having $M$ losses in the ancillary mode, for example, with $k_3=M$, would introduce an $\overline{X}$ error to the ancillary mode. This error would propagate a phase error of $\theta = -\frac{\pi}{N}$ to the input mode, thereby flipping the outcome of the $\overline{X}$ measurement there and introducing a logical error to the output. Consequently, the circuit can only tolerate up to $k<M$ losses. Therefore, choosing $N>M$ would not increase the tolerance of the circuit but would decrease the tolerance against phase errors in the input mode. A similar argument applies when $M>N$: Increasing $M$ decreases the tolerance against phase errors in the ancillary mode, while the circuit remains fault tolerant against only $N$ losses. Thus, we conclude that the optimal choice is to set $M=N$.

\subsection{Hybrid EC gadget}
\label{app:ft_hybrid}

\begin{figure}[h]
	\centering
	\includegraphics[trim=0mm 125mm 90mm 0mm, clip, width=0.5\columnwidth]{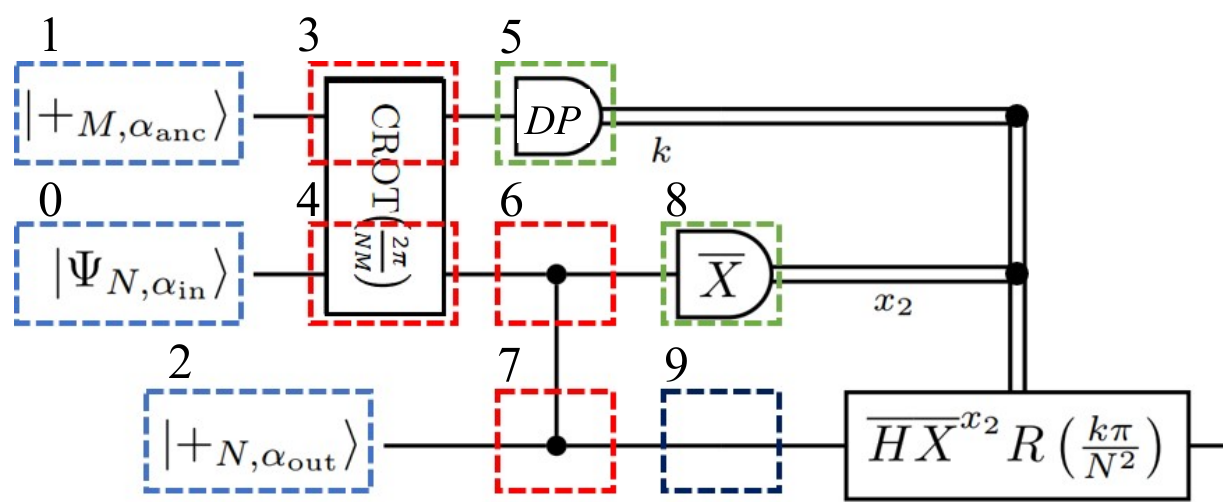}
	\caption{Fault locations in the hybrid EC gadget.}
	\label{fig:hybridFT}
\end{figure}

There are nine locations in the hybrid EC gadget, as shown in Fig.~\ref{fig:hybridFT}, labeled $i\in\{1,\dots,9\}$; again, the label $0$ is reserved for the input location. In this case, the top mode is the ancillary mode, the middle is the input mode, and the bottom is the output mode. As done earlier for the Knill situation, we can determine the number of losses and the magnitude of the phase errors for each mode before the final measurements:
\begin{align}
	\label{app:hybridEq}
	k^{(I)} &= k_0+k_4 + k_6 + k_8\,,\\
	\theta^{(I)} &= \theta_0+\theta_4 + \theta_6 + \theta_8 - \frac{2\pi k_1}{NM} -\frac{\pi k_2}{N^2}\,,\nonumber\\
	k^{(A)} &= k_1+k_3 + k_5\,,\nonumber\\
	\theta^{(A)} &= \theta_1 + \theta_3 + \theta_5 - \frac{2\pi k_0}{NM}\,,\nonumber\\
	k^{(O)} &= k_2 +k_7+ k_9,\nonumber\\
	\theta^{(O)} &= \theta_2 +\theta_7+ \theta_9 - \frac{\pi (k_0+k_4)}{N^2}.\nonumber
\end{align}
Following the earlier notation, we also have $r=\sum_{i=1}^9 k_i$, $\theta_f = \sum_{i=1}^9 \theta_i$, $\theta_s=-\frac{\pi k_0}{N^2}-\frac{2\pi k_0}{NM}$, and $\theta_r=-\frac{\pi(k_2+ k_4)}{N^2} - \frac{2\pi k_1}{NM}$.

Before we complete the proof, let us take a moment to discuss the value of $M$. Consider the following cases:
\begin{itemize}
	\item $2 \leq M \leq N$: When there are $\lceil M/2 \rceil$ or more losses in the ancilla mode ($k_1 \geq \lceil M/2 \rceil$), a $\overline{Z}$ error is propagated to the input mode as $|\theta^{(I)}| \geq \frac{\pi}{N}$. Therefore, the tolerance capacity of the circuit is limited by $\lceil M/2 \rceil$.
	\item $M>N$: Even in this case, the capacity of the circuit is still limited by $\lceil M/2 \rceil$. Therefore, given that $M>N$, to make the circuit fault tolerant against $N-1$ losses, we should choose $M \geq 2N-1$.
	\item $M=1$: In this case, the propagated phase error is $-\frac{2\pi k_1}{N}$. Regardless of the value of $k_1$, this error is equivalent to the identity on the code space.
\end{itemize} 
From these considerations, we observe that choosing $M = 1$ ensures that the circuit is resilient to losses in the ancilla mode and maximizes the tolerance against phase errors in the ancilla mode. We hence continue with the proof assuming $M = 1$.

Suppose $0\leq k_0+r < N$ and $|\theta_f + \theta_s + \theta_r |< \frac{\pi}{N}$. Here, $\theta_r$ can be set to $-\frac{\pi(k_2 + k_4)}{N^2}$ as a rotation of $\frac{2\pi k_1}{N}$ for integer $k_1$ is equivalent to the identity. It is evident that $k^{(n)} < N$ for each mode. For the input mode, we also have $|\theta^{(I)} |< \frac{\pi}{N}$ and thus the $\overline{X}$ measurement still gives the correct outcome. For the ancillary mode, the total phase error is $|\theta_1 + \theta_3 + \theta_5 |< \frac{\pi}{N}$. So, $k_0$ is still reliably detected. As a consequence, after applying the correction $R(\frac{\pi k_0}{N^2})$ to the output mode, the output state has the phase error of $|\theta_2+\theta_7 + \theta_9 - \frac{\pi k_4}{N^2}| < \frac{\pi}{N}$ and losses $k^{(O)}\leq r$. These are correctable errors and caused by faults inside the gadget. Therefore, we conclude that the hybrid EC gadget satisfies the EC-FT property.

\section{Simulation details}\label{app:Sim}

Here, we provide further details of the numerical simulation described in Sec.~\ref{sec:num_sim}.

\subsection{Mathematical recovery map}\label{app:Sim1}
In Sec.~\ref{sec:num_sim}, we mention the use of an ideal mathematical recovery map to separate the failure mode of the exRec from other situations. As the cat codes at hand are approximate codes, there is ambiguity in the choice of this mathematical recovery map---they differ in how the ``overlap" regions (nontrivial only for approximate codes) are attributed to different errors, and one can numerically find an \emph{optimal} recovery for a given error correction figure-of-merit. For computational simplicity, we elect to use an analytically defined recovery map, the transpose channel (also referred to as the Petz recovery \cite{Petz1986sufficient}), known to give near-optimal performance \cite{NgMandayam2010}. In any case, because these overlap regions are exponentially small as $\alpha$ grows, we do not expect significant differences in our conclusions even if we had used the optimal recovery here. 

\subsection{Average and entanglement fidelities}\label{app:Sim2}
In the main text, we said that we compute the ratio $R$ of the average infidelity and the benchmark infidelity for each instance of fault insertions in the exRec. Here, we explain how that $R$ is calculated. 

Rather than computing the average fidelity directly, which requires a Haar average over input states, we begin with the entanglement fidelity. For each instance of fault insertions, we compute the entanglement fidelity $\cF_\mathrm{ent}$ between that instance of faulty exRec and the ideal exRec,
\begin{equation}
	\cF_\mathrm{ent}\equiv \bra\Phi(\cI\otimes\cE){\left(\ketbra\Phi\right)}\ket\Phi.
\end{equation}
Here, $\ket\Phi$ is a maximally entangled state between the (two-dimensional) code space and an equal-dimensional reference space, $\cE$ is a (completely positive and trace-preserving) map that describes the action of the faulty exRec on the code space, while $\cI$ is the identity channel on the reference space. The entanglement fidelity quantifies how far the exRec is from the identity map---an exRec functions as the identity operation as long as the faults that occur in it cause only correctable errors---and the deviation from 1 indicates a failure of the exRec and a resultant logical error. We thus compute the entanglement \emph{in}fidelity, $\inF_\mathrm{ent}\equiv 1-\cF_\mathrm{ent}$, for each instance of fault insertions, and average that over the many runs. This many-runs average then tells us the failure rate of the exRec and hence the efficacy of the error correction scheme. 

Note that the entanglement fidelity is related to the average fidelity $\overline\cF$ of the output state of the exRec with the input state, with the average taken to be a Haar average over pure input states. Specifically, Ref.~\cite{Horodeckis1999general} showed that $\overline\cF=\frac{1}{d+1}(d\cF_\mathrm{ent}+1)$, where $d$ is the dimension of the reference space ($d=2$ here). This Haar average fidelity may be a more natural way of understanding the efficacy of this memory exRec, in its ability to preserve the stored state. The relation between entanglement and average fidelities tells us that the infidelities are in fact directly proportional:
\begin{equation}
	1-\overline\cF=\frac{d}{d+1} (1-\cF_\mathrm{ent})=\inF.
\end{equation}

We do a similar calculation using the entanglement fidelity for the benchmark situation, giving $\inFbm$ as the entanglement infidelity. Given that $d=2$ for both error-corrected (code space only) and benchmark cases, we get 
\begin{equation}
	R\equiv \frac{(1-\overline\cF)_\textrm{EC}}{(1-\overline\cF)\textrm{bm}}=\inF/\inFbm,
\end{equation}
equal to the ratio of the entanglement infidelities.

\end{document}